\documentclass[letterpaper,10pt]{scrartcl}
\usepackage[margin=1in]{geometry}
\usepackage[numbers,sort&compress]{natbib}
\usepackage{bm,epsfig,enumerate,tabularx,lmodern,fix-cm,mathtools,siunitx,listings,multirow,booktabs,rotating,floatrow,subfig,newfloat,abstract,authblk,url,blindtext,pdfpages,pgffor}

\usepackage[hang,flushmargin]{footmisc}

\begin{document}
	
\title{\LARGE Signatures of human impact on self-organized vegetation in the Horn of Africa}
\author[1]{Karna Gowda}
\author[2]{Sarah Iams}
\author[3]{Mary Silber\thanks{~~Correspondence to msilber@uchicago.edu}}

\affil[1]{\small Department of Engineering Sciences and Applied Mathematics, Northwestern University, Evanston, IL 60208, USA}
\affil[2]{\small Paulson School of Engineering and Applied Sciences, Harvard University, Cambridge, MA 02138, USA} 
\affil[3]{\small Committee on Computational and Applied Mathematics, and Department of Statistics, University of Chicago, Chicago, IL 60637, USA}
\date{}
\saythanks
\maketitle
\vspace{-25pt}
\begin{onecolabstract}
\noindent In many dryland environments, vegetation self-organizes into bands that can be clearly identified in remotely-sensed imagery. The status of individual bands can be tracked over time, allowing for a detailed remote analysis of how human populations affect the vital balance of dryland ecosystems. In this study, we characterize vegetation change in areas of the Horn of Africa where imagery taken in the early 1950s is available. We find that substantial change is associated with steep increases in human activity, which we infer primarily through the extent of road and dirt track development. A seemingly paradoxical signature of human impact appears as an increase in the widths of the vegetation bands, which effectively increases the extent of vegetation cover in many areas. We show that this widening occurs due to altered rates of vegetation colonization and mortality at the edges of the bands, and conjecture that such changes are driven by human-induced shifts in plant species composition. Our findings suggest signatures of human impact that may aid in identifying and monitoring vulnerable drylands in the Horn of Africa.
\vspace{25pt}
\end{onecolabstract}

Bands of vegetation separated by stretches of bare ground on gradually-sloping terrain are widespread in the drylands of Africa, North America, and Australia~\cite{Valentin:1999by,Deblauwe:2008if}. Vegetation bands often occur on pastoral lands~\cite{Gomes:2006vq,Deblauwe:2011ee}, and serve as crucial buffers against erosion~\cite{Valentin:1999gm, Ludwig:2005vi}. In some highly arid parts of the Horn of Africa (e.g., the Sool Plateau of Somalia), bands comprise the bulk of vegetation on the landscape. In general, dryland vegetation is susceptible to degradation due to overgrazing and changes in land use by human populations~\cite{Board:2005wo, Reynolds:2007bl}, both of which are relevant factors in the Horn of Africa~\cite{Gomes:2006vq}. Vegetation bands in this region are placed at additional risk by a recent multi-decadal decline in rainfall during the long rainy season (March--May)~\cite{Tierney:2015ct}, since large rainfall events during this season generate surface water runoff that is important for the maintenance of the bands~\cite{Aguiar:1999wk}. Vegetation bands are important hotspots of productivity in the changing and understudied drylands of the Horn of Africa~\cite{Durant:2012ey, Maestre:2012kj}, and understanding their response to human and climatic pressures can inform future conservation and relief efforts.

Mathematical models account for the emergence of vegetation bands via a self-organizing interaction between vegetation and water resources~\cite{Klausmeier:1999wo, vonHardenberg:2001bk, Rietkerk:2002ks, Gilad:2004bp}. Investigations of these models have sought signals of imminent catastrophic vegetation collapse in response to environmental change~\cite{Rietkerk:2004vq,Dakos:2011dp}. Most notably, the spacing between bands is predicted to increase in response to increases in aridity~\cite{Yizhaq:2005es, Sherratt:2007gq, vanderStelt:2012gs,Sherratt:2013ir,Siteur:2014jm} and local disturbances (e.g., grazing pressure)~\cite{Zelnik:2013iv}, resulting in overall reduced vegetation cover. In principle, reduced vegetation cover can also result from decreases in band width (the span of vegetation cover measured in the direction of local slope), but relatively little theoretical work has focused on this property of the bands. Previous empirical studies in Niger report reduced vegetation cover via decreased band width during multi-year periods of low rainfall~\cite{Valentin:1999gm,Wu:2000gu}. Reductions in cover in both studies were also associated with increased human activity, underscoring the importance of human impacts on vegetation change.

\begin{figure}[t]
\centering
\includegraphics[width=1\linewidth]{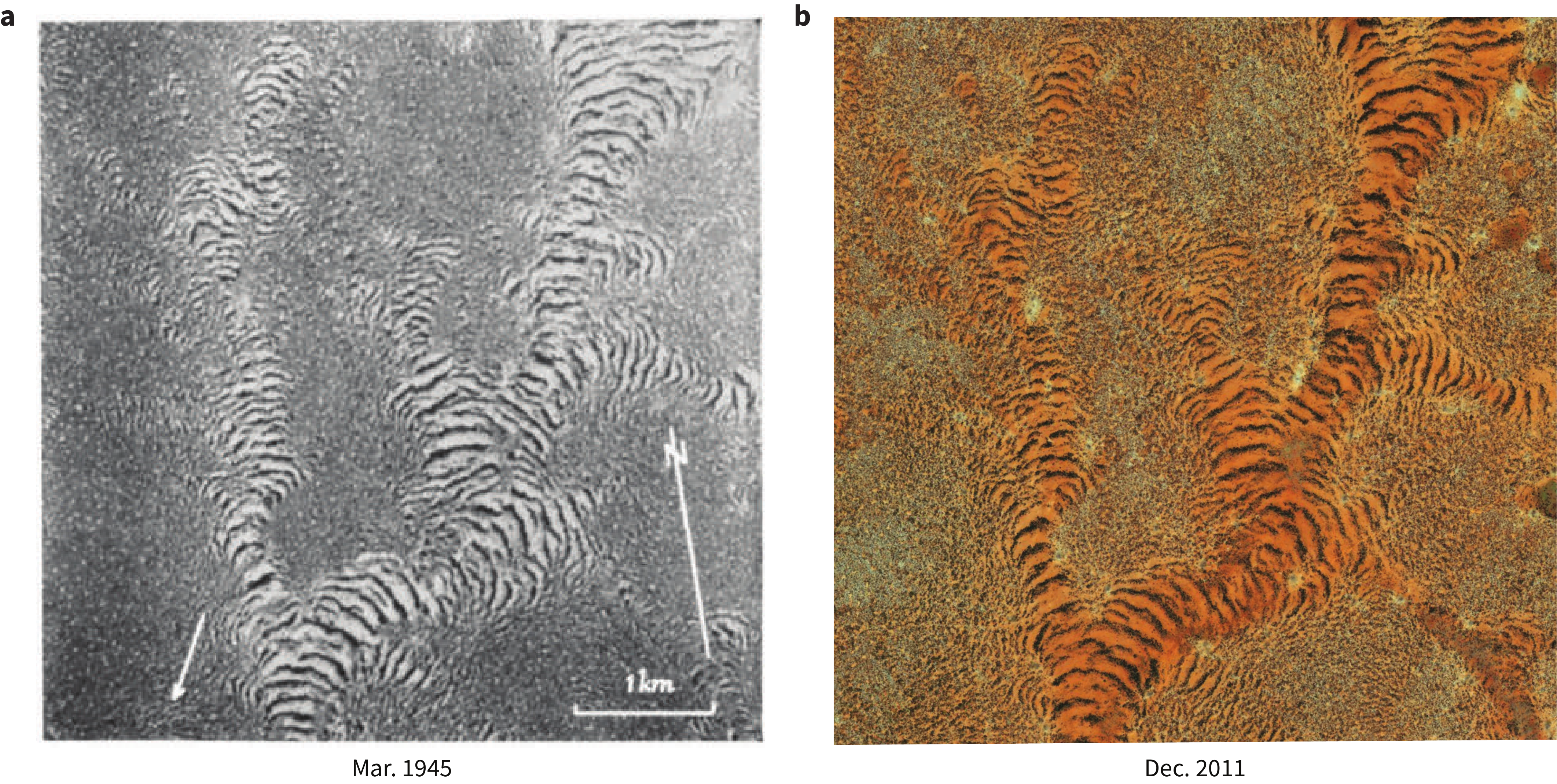}
\captionsetup{format=plain}
\caption{Aerial survey photographs taken over the Horn of Africa in the 1940s and 50s can be precisely referenced against modern imagery, enabling tracking of individual bands over a long period of time. (a) Aerial photograph adapted from plate 10 in~\cite{Macfadyen:1950vy}, with an arrow indicating the downslope direction. (b) Modern satellite image aligned to the photograph in (a) (\ang{7.85} N, \ang{47.41} E). Images \textcopyright~Royal Geographical Society, Google, DigitalGlobe.}
\label{fig:examples}
\end{figure}

\begin{figure*}[!!!t]
\centering
\includegraphics[width=1\linewidth]{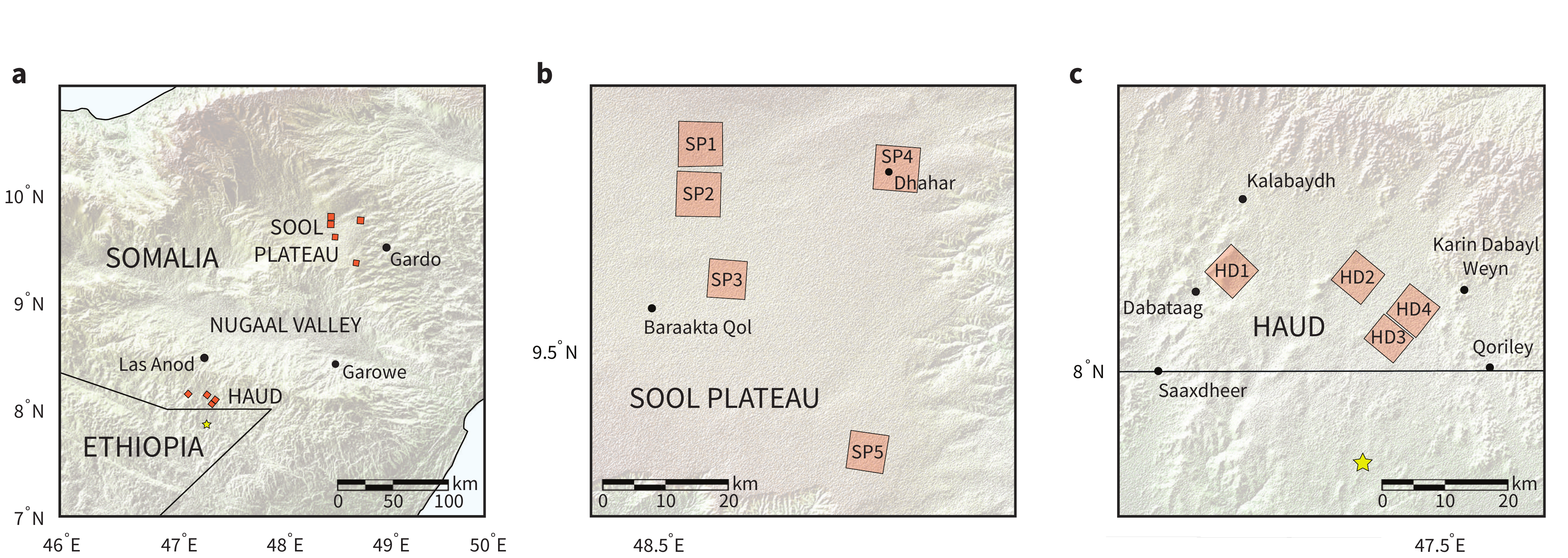}
\captionsetup{format=plain}
\caption{Areas studied in this investigation are defined by nine distinct aerial survey photographs taken in 1952. (a) Study areas are clustered in two regions of Somalia separated by the Nugaal Valley. (b) SP1--SP5 are located in the Sool Plateau pastoral region of Somalia. (c) HD1--HD4 are located in the Haud pastoral region of Somalia. Location of the example images shown in Figure~\ref{fig:examples} is indicated with a star. Each photograph covers an approximate area of 50 km$^2$. Relief maps rendered using the Shuttle Radar Topography Mission Global 1 arc second elevation dataset~\cite{Farr:2007ib} in MATLAB 2016b.}
\label{fig:sites}
\end{figure*}

Aerial survey photographs dating back to the 1940s and 50s were crucial to the initial discovery and characterization of vegetation bands in the Horn of Africa~\cite{Macfadyen:1950vy,Greenwood:1957ug} (Figure~\ref{fig:examples}). In the present study, we georeferenced high-resolution scans of survey photographs taken in 1952 and reconnaissance satellite imagery taken in 1967 against modern imagery to investigate the multidecadal dynamics of nearly 3,500 distinct vegetation bands. Our study areas cover approximately 450 km$^2$ of the Sool Plateau and Haud pastoral regions of Somalia (Figure~\ref{fig:sites}). Among these areas are locations that remain relatively unchanged, and also locations that have experienced a dramatic increase in human activity. All study areas experienced multi-year fluctuations in rainfall and a warming trend over the last half-century (Supplementary Information). We employed a systematic visual comparison of imagery to assess the extent of human activity and vegetation degradation, as well as automated transect-based measurements and comparative Fourier analysis to quantify key aspects of vegetation change.

\section*{Results}
\subsection*{Human activity and vegetation degradation}

\begin{figure}[!t]
\centering
\includegraphics[width=1\columnwidth]{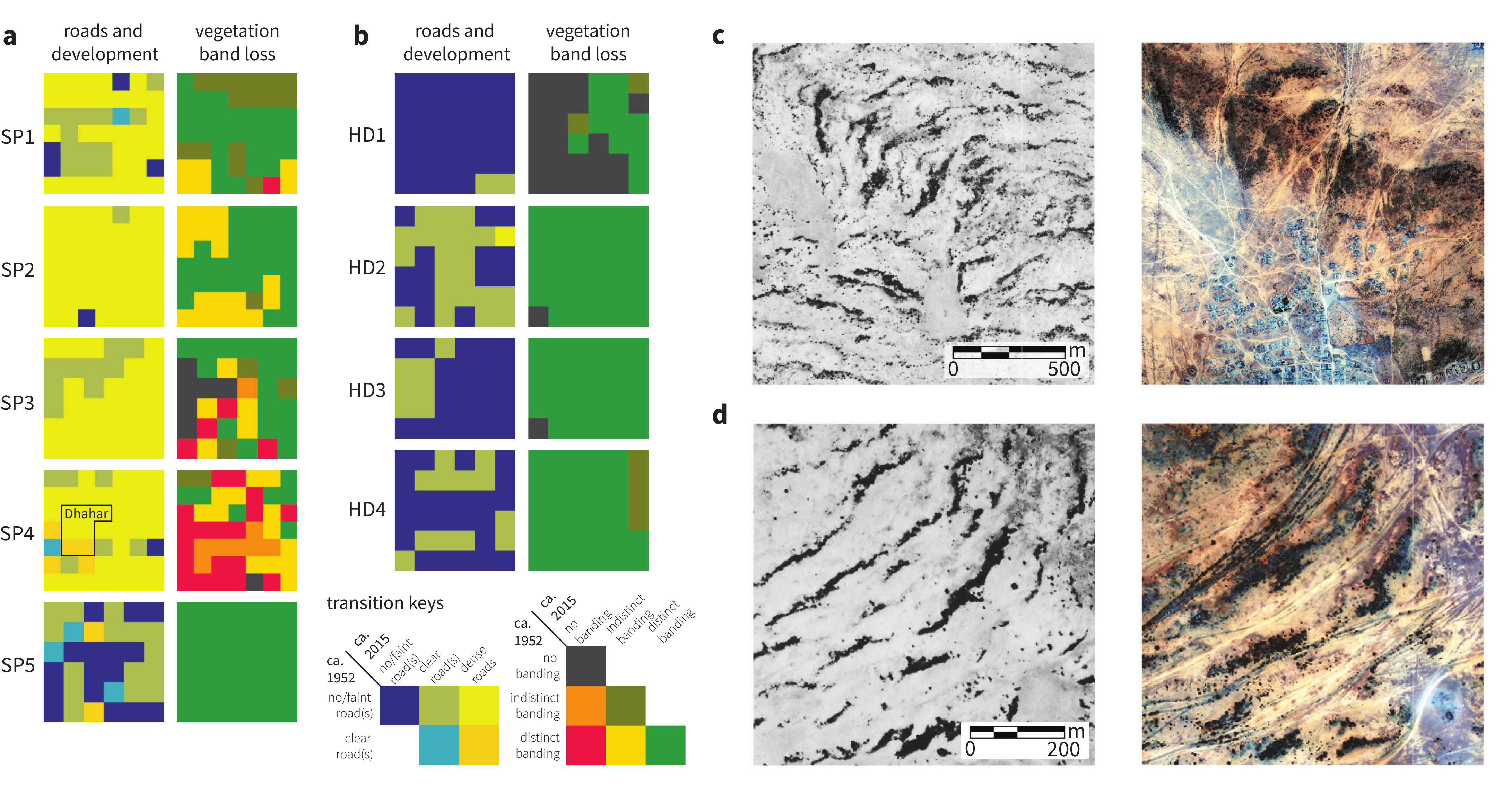}
\captionsetup{format=plain}
\caption{Vegetation loss occurs in areas with the steepest increases in human activity. The qualitative state of road and track cover, settlements, and vegetation banding was assessed visually in 1 km$^2$ boxes. Transitions between states observed in aerial photography and recent satellite imagery are shown here for all study areas. (a) SP1--SP4 in the Sool Plateau show a high degree of road and track development, and a moderate to high degree of band loss. A large settlement (Dhahar) developed within SP4, and is indicated with a black border. SP5 saw little increase in road cover, and no band loss or degradation was observed. (b) Areas in the Haud (HD1--HD4) show only a small increase of road and track development, and no substantial band loss or degradation is observed. (c) An example of band loss and degradation due to land development in SP4 (\ang{9.76} N, \ang{48.82} E; left: 02/22/1952, grayscale; right: 08/16/2016, RGB). (d) An example of band degradation amid dense track cover in SP3 (\ang{9.58} N, \ang{48.57} E; left: 11/29/1952, grayscale; right: 12/03/2011, RGB). Images courtesy of the Bodleian Library and the DigitalGlobe Foundation.}
\label{fig:vis_analysis}
\end{figure}

We assessed changes in human activity and vegetation banding over time through a systematic visual comparison of British Royal Air Force (R.A.F.) aerial survey photography taken in 1952 and recent satellite imagery (Methods, Supplementary Information). We found that band degradation ranged from partial to complete band loss in areas with the steepest increases in human activity, while bands in the other areas remain largely unchanged (Figure~\ref{fig:vis_analysis}). Roads and dirt tracks can be visually identified in both the aerial photographs and the satellite imagery, and their presence and qualitative appearance served as our primary proxy for inferring the extent of human activity (Supplementary Figure S4). We defined degradation in this study as either the breakdown in regularity or the disappearance of banding (Supplementary Figure S5).

Substantial road and track development occurred in much of the Sool Plateau, with most areas (SP1--SP4) transitioning from having either no roads or faint roads in 1952 to having roads or tracks that densely cover the landscape in the modern images (Figure~\ref{fig:vis_analysis}a). The settlement Dhahar was founded within SP4 after the 1952 photograph, and now supports a population of approximately 13,000 (Figure~\ref{fig:vis_analysis}c). We observed much less road and track development in Sool Plateau area SP5 and in all Haud areas (HD1--HD4) (Figure~\ref{fig:vis_analysis}b). At many sites within the Haud, human-made structures visible in the 1952 images seem to persist into the current decade, suggesting no major change in land use over the intervening time (Supplementary Figure S6a).

Vegetation degradation is prevalent in the human-impacted areas SP1--SP4 (Figure~\ref{fig:vis_analysis}a). Bands have disappeared entirely from the landscape in large parts of SP3 and SP4. Only part of the band loss in these areas seems directly related to clearing for land development, since loss also occurs in areas without human-made structures. In SP1--SP4, dense tracks often appear between bands (Figure~\ref{fig:vis_analysis}d). Frequently we observed vegetation growing within roads and tracks, which suggests that these structures likely disrupt the flow of water on the landscape. In SP5 and HD1--HD4, individual bands often remain identifiable after six decades based on visible details of their morphology, and we observed no substantial band degradation.

\subsection*{Band widening in human-impacted areas}
\begin{table}[t]
\centering \footnotesize
\captionsetup{format=plain}
\caption{Typical slopes, ratios of band widths between new and old imagery, and band migration rates in each study area. Ranges shown are from the 25th to 75th percentiles. The width ratios ($R$) and upslope colonization and retreat rates, $M_c$ and $M_r$ respectively, were measured between 1952 and c. 2010 (images used for analysis indicated in Supplementary Table S1). Significance of correlations was assessed using a one-tailed $t$-test corrected for spatial autocorrelation~\cite{Dutilleul:1993jn}, and $p$ values, $t$ values and degrees of freedom are given.}\label{tab:band_props}
\setlength{\tabcolsep}{2pt}
\begin{tabular}{cccccccc}

\toprule
\multirow{2}{*}{Area} & Slope (\%) & Width ratio & \multicolumn{2}{c}{Migration (m/yr)} & \multirow{2}{*}{corr($S$, $R$) ($p$, $t$, df)} & \multirow{2}{*}{corr($S$,$M_c$) ($p$, $t$, df)} & \multirow{2}{*}{corr($S$,$M_r$) ($p$, $t$, df)} \\
                      & $S$        & $R$         & $M_c$             & $M_r$            &                                       &                                        &                                        \\ \midrule
SP1                   & 0.3--0.4   & 0.95--1.5   & 0.26--0.68        & 0.12--0.48       & -0.21 ($< 0.005$, 8.7, 198)   & -0.19 (0.01, 7.8, 199)         & 0.01 (0.91, 0.0, 219)                            \\
SP2                   & 0.1--0.3   & 1.3--2.2    & 0.57--1.2         & 0.20--0.55       & -0.12 ($< 0.005$, 9.5, 666)   & -0.1 (0.01, 6.1, 659)          & 0.03 (0.43, 0.6, 629)                            \\
SP3                   & 0.1--0.3   & 1.0--1.7    & 0.23--0.79        & 0.08--0.51       & -0.1 (0.13, 2.4, 237)                & -0.09 (0.16, 2.0, 237)                & -0.05 (0.47, 0.5, 219)                           \\
SP4                   & 0.1--0.2   & 1.2--1.9    & 0.33--0.93        & 0.08--0.41       & -0.12 (0.14, 2.2, 152)               & -0.12 (0.17, 1.9, 136)                & 0.00 (0.97, 0.0, 137)                            \\
SP5                   & 0.2--0.4   & 0.74--1.1   & 0.17--0.38        & 0.20--0.43       & -0.13 (0.01, 6.4, 346)        & -0.14 ($< 0.005$, 9.1, 435)    & -0.06 (0.21, 1.6, 492)                           \\ \midrule
HD1                   & 0.4--0.6   & 0.93--1.4   & 0.22--0.55        & 0.07--0.39       & -0.13 (0.16, 2.0, 123)               & -0.32 ($< 0.005$, 16, 141)     & -0.24 (0.01, 8.1, 136)                           \\
HD2                   & 0.3--0.5   & 0.85--1.2  & 0.25--0.47        & 0.21--0.46       & 0.12 (0.06, 4.1, 289)                 & -0.06 (0.31, 1.1, 324)                           & -0.16 ($< 0.005$, 8.5, 314)               \\
HD3                   & 0.3--0.5   & 0.87--1.2   & 0.26--0.48        & 0.22--0.46       & 0.03 (0.65, 0.2, 321)                & -0.14 (0.01, 7.9, 420)        & -0.18 ($< 0.005$, 13, 386)                      \\
HD4                   & 0.4--0.5   & 0.85--1.2   & 0.19--0.41        & 0.15--0.39       & -0.07 (0.18, 1.8, 340)               & -0.14 (0.01, 16, 273)        & -0.16 (0.01, 7.1, 280)
\\ \bottomrule
\end{tabular}
\end{table}

\begin{figure}[t]
\centering
\includegraphics[width=1\linewidth]{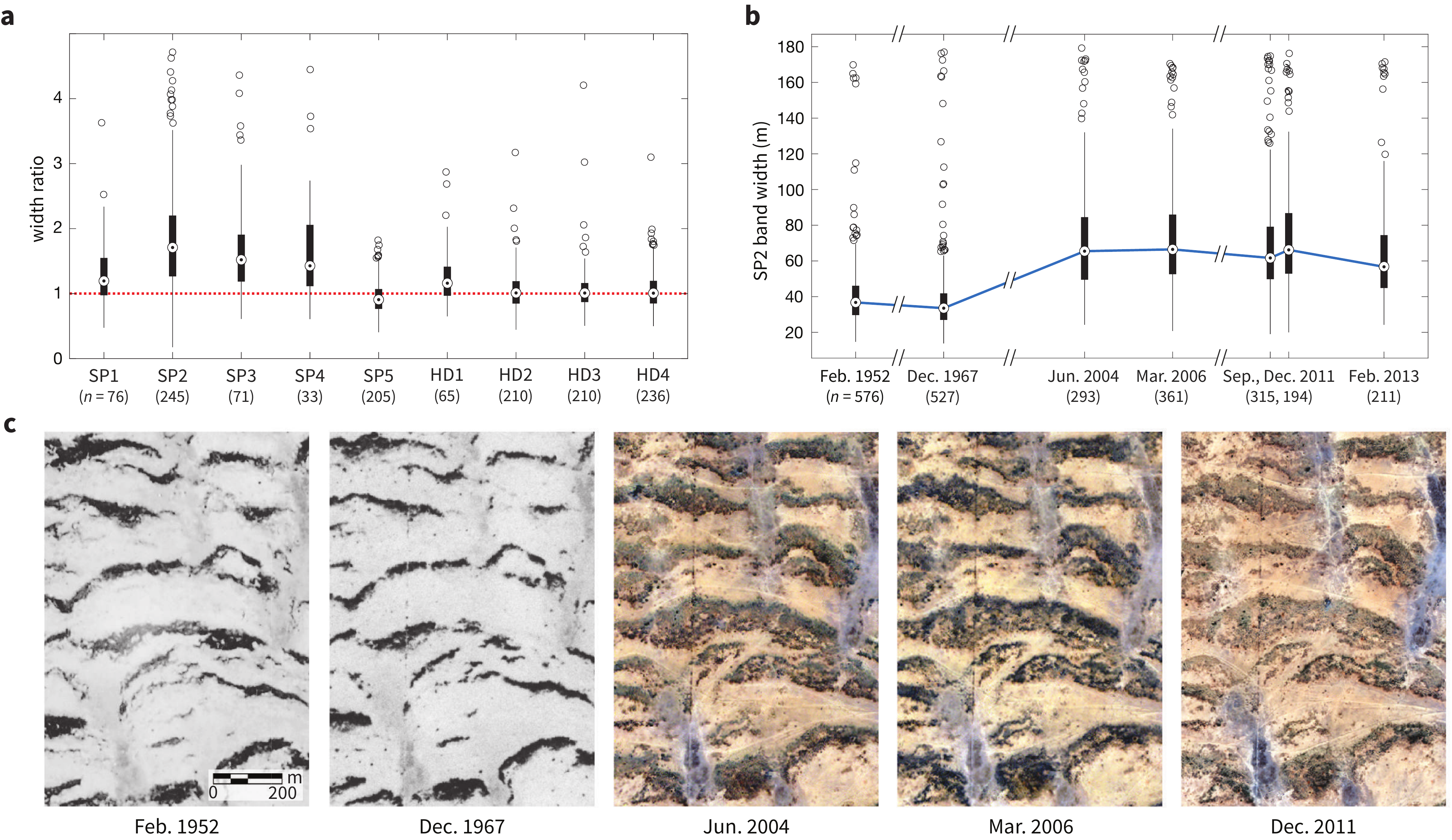}
\captionsetup{format=plain}
\caption{Bands widen appreciably in the direction of slope in the most heavily human-impacted areas, SP1--SP4. (a) The ratio of widths measured in a recent image to widths measured in a 1952 photograph are shown for all study areas. (b) The band widths measured in SP2 are shown at seven points in time. Widths change little between 1952 and 1967, and nearly double between 1967 and 2004. (c) An example of band widening in SP2 (\ang{9.73} N, \ang{48.55} E). Images courtesy of the Bodleian Library, U.S. Geological Survey, and DigitalGlobe Foundation.}
\label{fig:widening}
\end{figure}

We quantified aspects of vegetation dynamics in all study areas using automated transect measurements of individual bands (Methods, Supplementary Information). Most notably, we found that bands have widened appreciably in the direction of slope in the areas SP1--SP4, while band widths remained approximately constant in the other study areas (Figure~\ref{fig:widening}, Table~\ref{tab:band_props}). This widening, which results in increased vegetation cover, is surprising in light of the apparent steep increases in human activity in these areas.

We computed the ratio of band widths measured in recent imagery to the widths in 1952. The median ratio among bands in each area exceeded 1.2 in SP1--SP4 (Figure~\ref{fig:widening}a). The most substantial widening occurred at SP2, where the median ratio is 1.7. We measured band widths at SP2 using additional images taken in 1967, 2004, 2006, 2011, and 2013. We found that widths did not change between 1952 and 1967, and then nearly doubled between 1967 and 2004 (Figure~\ref{fig:widening}b,c). From 2004 onward, median band width held approximately constant. Analyses over multiple time points at SP1, SP3, and SP4 showed a similar pattern (Supplementary Figure S8). Since recent images were taken in a variety of seasonal and rainfall-history conditions (Supplementary Figure S1), we conclude that the widening observed in SP1--SP4 is not an artifact of seasonality. Moreover, we do not observe widening at nearby area SP5, which strongly suggests that non-climatic factors have driven the apparent changes in SP1--SP4.

\begin{figure}[t]
\centering
\includegraphics[width=1\linewidth]{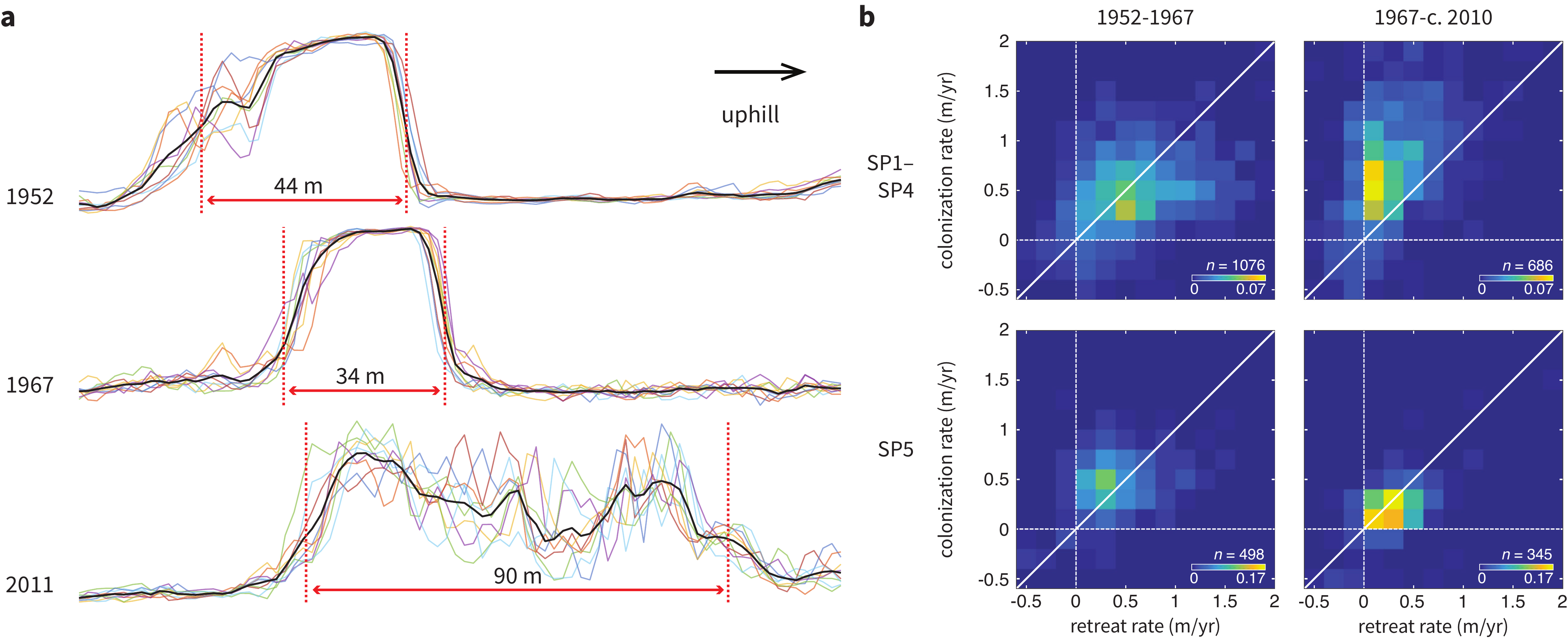}
\captionsetup{format=plain}
\caption{Bands widen in areas SP1--SP4 due to increased upslope edge migration (colonization) and decreased downslope edge migration (retreat). (a) Example image intensity profiles along a transect in SP2. Profiles along multiple parallel transects are shown in color, and the mean profile is shown in black. The estimated band edges are indicated in red. Intensity values are normalized to the same scale for comparison. (b) Bivariate distribution of front and back migration rates shown for the periods 1952--1967 (first column) and 1967--c. 2010 (second column), with study areas SP1--SP4 (first row) compared in aggregate with SP5 alone (second row). Band widening in SP1--SP4 results from front migration rates increasing and back migration rates decreasing during the period 1967--c. 2010.}
\label{fig:migration}
\end{figure}

Vegetation bands in Africa and North America are reported to migrate uphill over time due to vegetation colonization at the upslope edge of the band and mortality-driven retreat at the downslope edge~\cite{Valentin:1999by, Deblauwe:2012ih}. In all our study areas, we similarly observed that bands gradually migrate uphill over six decades (Table~\ref{tab:band_props}). In addition, we found that bands widen at SP1--SP4 due to increased uphill migration rates at the upslope edges of the bands (colonization rate) and to decreased rates at the downslope edges (retreat rate) (Figure~\ref{fig:migration}). Between 1952 and 1967, both colonization and retreat rates are comparable in all Sool Plateau study areas, resulting in band widths that remain unchanged over this period. Between 1967 and c. 2010, colonization rates increase and retreat rates decrease in SP1--SP4, resulting in band widening. In contrast, colonization and retreat rates both decrease in SP5 by the same factor during this period, resulting in slower migration and no widening in this study area. 

We found evidence that migration rates and band width ratios vary inversely with local slopes, which are typically shallow and range between 0.1--0.6\% (Table~\ref{tab:band_props}). Colonization rates are negatively correlated with slope in a majority of study areas ($p \leq 0.01$), including the areas in the Sool Plateau where width ratios are also negatively correlated with slope ($p \leq 0.01$). Retreat rates are only negatively correlated with slope in the Haud study areas ($p \leq 0.01$). Since widening in SP1--SP4 occurred due to increased colonization rates, this suggests that bands tend to widen to a greater degree on shallower slopes due to faster rates of colonization. An inverse relationship between slope and migration rate seems to contradict previous investigations of a mathematical model for vegetation banding, which indicate either a negligible~\cite{Sherratt:2005hy} or an increasing relationship~\cite{Sherratt:2010ga}. Moreover, a positive relationship between migration and slope is expected in the limit of vanishing slope, since the migration rate must approach zero as the anisotropy induced by the slope vanishes. The bands in regions of study may rely critically on the slopes being above some threshold and this could explain why we do not observe positive relationships between slope and migration.

\subsection*{Comparison with previous theoretical predictions}
Theoretical investigations of vegetation band response to environmental pressure have focused on the characteristic spacing between bands (band wavelength) because this property is observable in remotely-sensed imagery. Idealized partial differential equation models, which account for water-biomass feedbacks and transport, predict that band wavelength should increase in response to sufficient increases in environmental pressure~\cite{Yizhaq:2005es, Sherratt:2007gq, vanderStelt:2012gs,Sherratt:2013ir,Siteur:2014jm,Zelnik:2013iv}. We quantified band wavelength change in all study areas using the Fourier window method developed by Penny \textit{et al.}~\cite{Penny:2013ed} (Methods, Supplementary Information). In areas where bands have not completely disappeared, we found that changes in wavelength are imperceptible (Supplementary Information). In parts of the human-impacted area SP4, bands look to have degraded without an apparent change in wavelength (Supplementary Figure S6b).

Band wavelength is also predicted by models to vary with local slope, though the nature of this relationship can depend on model parameters and the history of the state~\cite{Sherratt:2005hy, Sherratt:2015cq}. We found a significant correlation at the 5\% level between wavelength and slope in only SP3 ($r = -0.34$, $p = 0.04$) (Supplementary Table S2). The (negative) sign of this correlation agrees with empirical findings in other investigations~\cite{Penny:2013ed, Deblauwe:2011ee, Deblauwe:2012ih}.

\section*{Discussion}
In this study, we observed two distinct environmental outcomes that depend on the level of apparent human activity. In the areas with only modest increases in activity, remarkably little about the vegetation has changed. In the areas with significant increases in activity, the vegetation has degraded substantially, with bands often disappearing entirely from the landscape. Vegetation loss is not limited to land cleared for development, since we also observed loss in areas with few human-made structures. This suggests more subtle forms of relevant human impact. Roads and dirt tracks now densely cover the landscapes of many of the study areas. This may be indicative of overgrazing or biomass harvesting, both of which are known issues in the area~\cite{Oduori:2003uy}. The roads themselves likely also affect the flow and availability of water to the vegetation bands~\cite{Hemming:1966fm}, which we observed most clearly in cases where vegetation grows within dirt tracks. Due to the faintness of the tracks in many instances, we assessed the extent of roads and dirt tracks visually. Developing automated image analysis approaches for detecting faint roads and dirt tracks would enable remote-sensing proxies for human activity that may be used to monitor the Horn of Africa and other pastoral lands.

In many study areas, we observed an increase in the widths of vegetation bands driven by altered rates of upslope and downslope edge migration. Widening occurs at Sool Plateau areas SP1--SP4, and does not occur in nearby area SP5, suggesting non-climatic causes. Since widening occurred in conjunction with apparent increases in human activity, we investigated how widening may be linked to human impacts by numerically simulating a well-studied dryland vegetation model under changes in parameters associated with a possible shift from trees to grasses~\cite{Klausmeier:1999wo}. Although we could not infer plant composition changes from the satellite imagery, it has been documented that \textit{Acacia} woodcutting for charcoal production is widespread in the Sool Plateau~\cite{Oduori:2003uy}, and has likely caused a decrease in woody biomass within the vegetation bands in many areas. A shift in composition that decreases woody biomass and increases grass biomass plausibly increases the overall transpiration rate, biomass yield per unit water, dispersal rate, and mortality rate of the vegetation. We found that individually increasing transpiration, yield, and dispersal rate parameters in the model cause band widening, while increasing mortality rate had the opposite effect (Supplementary Figure S11). We also found that widening can be achieved through a simultaneous increase in transpiration, biomass yield, dispersal, and mortality rate parameters (Supplementary Figure S11b). We conclude that a shift in species composition is a viable explanation for the changes we have observed in the Sool Plateau. 

Band widening in the human-impacted study areas causes local increases in vegetation cover. This is seemingly at odds with our observations of band degradation within the same areas, and also with previous studies of vegetation bands in Niger reporting diminished vegetation cover accompanying greater human activity~\cite{Valentin:1999gm,Wu:2000gu}. Our investigation of a dryland vegetation model posits that woodcutting, a relevant impact of human activity, can induce band widening. A recent theoretical study suggests also that increased pastoral grazing pressure can cause band widening and increased vegetation cover~\cite{SIERO201864}. Additional remote-sensing studies could clarify the relationship between human activity and vegetation band widening. In the present study, we analyzed only a small fraction of the 1950s aerial survey photography archive. A future study making use of randomly-sampled photographs over a larger area could better characterize the prevalence of vegetation band widening, and the frequency of its association with increased human activity. A well-resolved time series of images, e.g., from the Landsat or Sentinel satellites, could be used for a phenological analysis of vegetation that permits a spatial classification of plant functional type within bands. Such information could be used to determine how plant composition varies between bands that have widened and those that have not. Future monitoring of the region could help to determine whether band widening is ultimately a transient phenomenon, and to characterize the relationship between band widening and vegetation resilience.

We have provided new evidence suggesting that human impacts modulate the width of vegetation bands, and we argue that this vegetation property represents an underutilized window into the response of dryland vegetation to environmental pressure. In recent years, the spacing between bands (band wavelength) has been a prime focus of theoretical and empirical investigations of vegetation pattern resilience. As we and others~\cite{Penny:2013ed} have found, the spacing between bands in nature is often irregular, making both the notion of wavelength and its measurement imprecise. In models of vegetation patterning, it is common for a range of band wavelengths to be stable over a range of environmental conditions, making wavelength changes in model scenarios of environmental change history-dependent and discontinuous~\cite{Sherratt:2007gq,vanderStelt:2012gs,Sherratt:2013ir,Siteur:2014jf}. In contrast, band widths are straightforward to measure in remotely-sensed imagery, and our model investigation suggests that band widths change continuously in response to parameter variation (Supplementary Figure S11). Future theoretical and empirical investigation of this vegetation property will be important to establishing its utility to dryland monitoring in the Horn of Africa.   

{\section*{Methods}
\subsection*{Regional information} We studied areas within the Sool Plateau and Haud pastoral regions of Somalia. Both regions are generally characterized by an arid climate (aridity index = 0.04--0.1)~\cite{Muchiri:2007ue}. Due to a lack of continuous rainfall station monitoring in and around our regions of study, we assessed the historical regional climate using 20th Century Reanalysis~\cite{Compo:2011iq} and the CPC/Famine Early Warning System Dekadal Estimates datasets (Supplementary Figure S1). Mean annual rainfall in both regions ranges between 100--300 mm. We found no evidence that rainfall conditions have changed in either region in recent decades, and we identified a warming trend in average yearly temperature of 1--2 $\si{\degree}$C over the last half-century.

Regional soils are claylike and prone to crust formation, resulting in low permeability and surface water runoff following high-intensity rainfall~\cite{Hemming:1965tr,Oduori:2003uy}. Hemming found that soils are wetter beneath bands in the Haud, indicating greater soil permeability in vegetated areas~\cite{Hemming:1965tr}. Vegetation bands in both regions are dominated by \textit{Andropogon kelleri} grasses~\cite{Macfadyen:1950vy,Hemming:1965tr}. Bands also contain a mix of trees and shrubs, most notably \textit{Acacia bussei}. In recent decades, \textit{Acacia bussei} has diminished in abundance in the Sool Plateau due to cutting for charcoal production~\cite{Oduori:2003uy}. Disruption of traditional grazing patterns has resulted in overgrazing in many areas of the Sool Plateau, including Dhahar (SP4)~\cite{Oduori:2003uy}.

\subsection*{Data} We studied approximately 260 km$^2$ of imagery within the Sool Plateau and 200 km$^2$ of imagery within the Haud (Figure~\ref{fig:sites}). Study areas were chosen based on a combination of factors. In particular, we wished to include areas with different development and degradation outcomes, areas with recorded soil and floristic information based on field studies, areas in geographically distinct regions, and areas featuring well-defined banding. Study area boundaries are defined by our choice of British Royal Air Force (R.A.F.) aerial survey photography, which comprise our earliest image datasets. Aerial survey photographs were taken in 1951--52 over broad areas of present-day Somalia, and specific photographs were scanned on request by the Bodleian Library at the University of Oxford. We also studied declassified reconnaissance satellite imagery taken in 1967, and DigitalGlobe imagery for dates spanning 2004--2016. Resolution of imagery used in this study ranges between 1.4--2.4 m/pixel. Satellite images taken between 2004--2016 containing red and near-infrared channels were used to compute a Soil-adjusted Vegetation Index~\cite{Huete:1988ez}. We manually georeferenced R.A.F. scans and the 1967 reconnaissance image using visually identified control points. We estimated georeferencing error to be approximately 1--2 pixels.

We estimated local gradient within our study areas using the Shuttle Radar Topography Mission Global 1 arc second elevation dataset~\cite{Farr:2007ib}. Because of the noise characteristics of the dataset and the low relief of our study areas, we used a second-order finite difference operator with noise-suppressing properties to estimate gradient and slope~\cite{NRIGO} (Supplementary Figure S2).

\subsection*{Visual comparison} We assessed changes in our study areas over time through a systematic visual comparison of imagery. We developed a graphical user interface in MATLAB for comparing images (Supplementary Figure S3). For each area, we split both R.A.F. scans and recent imagery into 1 km $\times$ 1 km boxes, and evaluated qualitative features within these boxes. We evaluated the extent of roads and dirt tracks, which served as our primary proxy for human pressure (Supplementary Figure S4). We also evaluated the extent of banding and defined degradation in this context as the breakdown in regularity or disappearance of banding between the earliest and most recent images (Supplementary Figure S5).

\subsection*{Automated transect measurements} We quantified band widening and migration using grayscale image intensity profiles gathered along transects drawn through the bands. We used the same transects for multiple images in the same study area. We fit a top hat function to each intensity profile to extract band width (Supplementary Figure S7). Intensity measurements were gathered along multiple parallel transects for each band, and data points with high variance in measured widths were discarded.

\subsection*{Fourier analysis} We measured changes in band wavelength using the Fourier window method by Penny \textit{et al.}~\cite{Penny:2013ed}. The method measures band wavelength and orientation in a sliding window using a 2D FFT, and computes a uniqueness metric based on the unimodality of the radially-binned power spectrum. We discarded data points which correspond to sites without banding using a manually-drawn mask. We additionally discarded data points with uniqueness values below a threshold. 

\subsection*{Model simulations} We simulated the model by Klausmeier~\cite{Klausmeier:1999wo}, a conceptual model describing biomass-water interactions in dryland environments, in one spatial dimension. We estimated the sensitivity of band width to parameter changes. We obtained the initial parameter set from the values and ranges given in~\cite{Klausmeier:1999wo}. Parameters stated in~\cite{Klausmeier:1999wo} to differ between grasses and trees are set at intermediate values so that the spatial scale of banding resembles those in our regions of study. The time scale of migration was similarly tuned using the downhill water flow rate parameter. We simulated the model using the exponential time differencing fourth-order Runge-Kutta pseudospectral scheme~\cite{Kassam:2005jv}.

\subsection*{Statistics} We assessed the significance of correlations between quantities computed from transect and Fourier analyses using a one-tailed paired $t$-test corrected for spatially autocorrelated data~\cite{Dutilleul:1993jn}, implemented in SpatialPack for R~\cite{Osorio:2016uy}.

\subsection*{Data availability} Data and code supporting the findings of this study are available at https://doi.org/10.21985/N2GQ1G.

{\bibliography{library}}
\bibliographystyle{naturemag}
\vspace{10pt} 
\noindent\textbf{Acknowledgements}\\
We are grateful for imagery grants from the DigitalGlobe Foundation, and to Michael Athanson and the Bodleian Library at the University of Oxford for locating and scanning aerial photographs. We thank Stefano Allesina, Punit Gandhi, Eric Siero, and Lucien Werner for helpful discussions, and Kelsey Rydland for ArcGIS support. Research was supported in part by NSF DMS-1517416, the NSF RTG in Quantitative Biological Modeling (DMS-1547394), and the NSF Math and Climate Research Network (DMS-0940262).
\\\\
\textbf{Author Contributions}\\
K.G., S.I., and M.S. designed research; K.G. performed research with input from S.I. and M.S.; K.G. analyzed data; K.G. wrote the paper with contributions and feedback from S.I. and M.S.}
\\\\
\textbf{Additional Information}\\
\textbf{Competing financial interests:} The authors declare no competing financial interests.

\clearpage
\newpage


\section*{Supplementary material}

# Supplementary information
## Signatures of human impact on self-organized vegetation in the Horn of Africa

Karna Gowda[1], Sarah Iams[2], and Mary Silber[*3]

[1]Department of Engineering Sciences and Applied Mathematics, Northwestern University, Evanston, IL 60208, USA
[2]Paulson School of Engineering and Applied Sciences, Harvard University, Cambridge, MA 02138, USA
[3]Committee on Computational and Applied Mathematics, and Department of Statistics, University of Chicago, Chicago, IL 60637, USA

# Contents

---

[*]msilber@uchicago.edu

# S1 Regional information

We studied imagery in areas located within the Sool Plateau and Haud pastoral regions of Somalia. Sool Plateau study areas are located approximately 50 km west of Gardo, and Haud areas are located approximately 40 km south of Las Anod. Areas were chosen for this study based on a combination of factors; in particular, we wished to include areas with different development and degradation outcomes, areas with recorded soil and floristic information based on field studies, areas in geographically distinct regions, and areas featuring well-defined banding. Detailed information about areas and imagery is given in Section S2.1 and Table S1.

## S1.1 Climate

Both Sool Plateau and Haud pastoral regions are characterized by an arid climate (aridity index = 0.04-0.1) [1]. Rainfall in Somalia is bimodally distributed between the Gu season, spanning Apr.-May, and the Deyr season, spanning Oct.-Nov. Separating the rainy seasons are two dry seasons, Xagaa (Jun.-Sept.) and Jilaal (Dec.-Mar.). Deyr rainfall events are typically shorter and less significant than those of the Gu. The Jilaal season is typically the hottest and driest time of year.

Due to a lack of continuous rainfall station monitoring in and around our regions of study, we assessed the historical regional climate using climate reanalysis and remotely-sensed rainfall estimation datasets. The 20th Century Reanalysis (V2c) dataset assimilates surface pressure observations, sea-surface temperature, and sea ice extent into a global climate model to obtain a reconstruction of Earth's climate spanning 1871-2011 [2]. The V2c dataset is available at 6-hour temporal and $2°$ spatial resolution. The coarse spatial resolution of the data prevents us from distinguishing the Sool Plateau and the Haud regions. Uncertainty estimates can be derived from 56 replicate model simulations. The CPC/Famine Early Warning System Dekadal Estimates (RFEv2) dataset uses satellite microwave sensing and ground station observations to estimate total rainfall over the African continent for dates spanning 2000 to present. RFEv2 data is available at daily intervals and $0.25°$ spatial resolution.

To assess the rainfall conditions surrounding our imagery datasets, we obtained annual total rainfall estimates from the V2c and RFEv2 datasets (Figure S1a-d). In the absence of ground confirmation, we exercise caution in interpreting the V2c estimates for the 1940s-60s, and conclude only that there is no evidence that rainfall conditions have improved in either region in recent decades. We speculate that rainfall conditions surrounding the 1952 and 1967 datasets were above average. We also speculate that conditions have either declined or reverted to a regional mean in recent decades.

We assessed rainfall conditions for the recent imagery in greater detail by calculating seasonal rainfall totals from the RFEv2 dataset (Figure S1c-d). In the Sool Plateau, the images used in this study were taken in a variety of seasons and rainfall history conditions. The 2004 image was taken shortly after the return of rains that followed a very severe multi-year drought. The 2006, 2011, and 2013 images were taken amid more typical rainfall conditions. The 2016 image was taken during a period of severe drought in northeastern Somalia, which is ongoing at the time of writing. In the Haud, the images used in this study were taken in 2012 and 2016, years with robust rainfall during the wet seasons.

We examined regional temperature history using surface temperature estimates from the V2c dataset. We computed the average yearly temperature, defined as a yearly average over the daily midpoint between minimum and maximum temperatures (Figure S1e). We identified a distinct linear warming trend between 1960 to the present of 1-2 °C.

## S1.2 Vegetation

Field investigations in the 1950s-60s found vegetation bands in the Sool Plateau and the Haud to be dominated by *Andropogon kelleri* grasses [3, 4]. The bands also usually contain a mix of trees and shrubs. Long-lived *Acacia bussei* trees often populate the bands. In many of our study areas, bands occur on relative low-ground (e.g., in channels), while a more uniform cover occurs on the surrounding higher ground. Greenwood attributes this difference to the clay content of the soil: higher-ground areas have a lower clay content, and thus can support a greater density of vegetation [5].

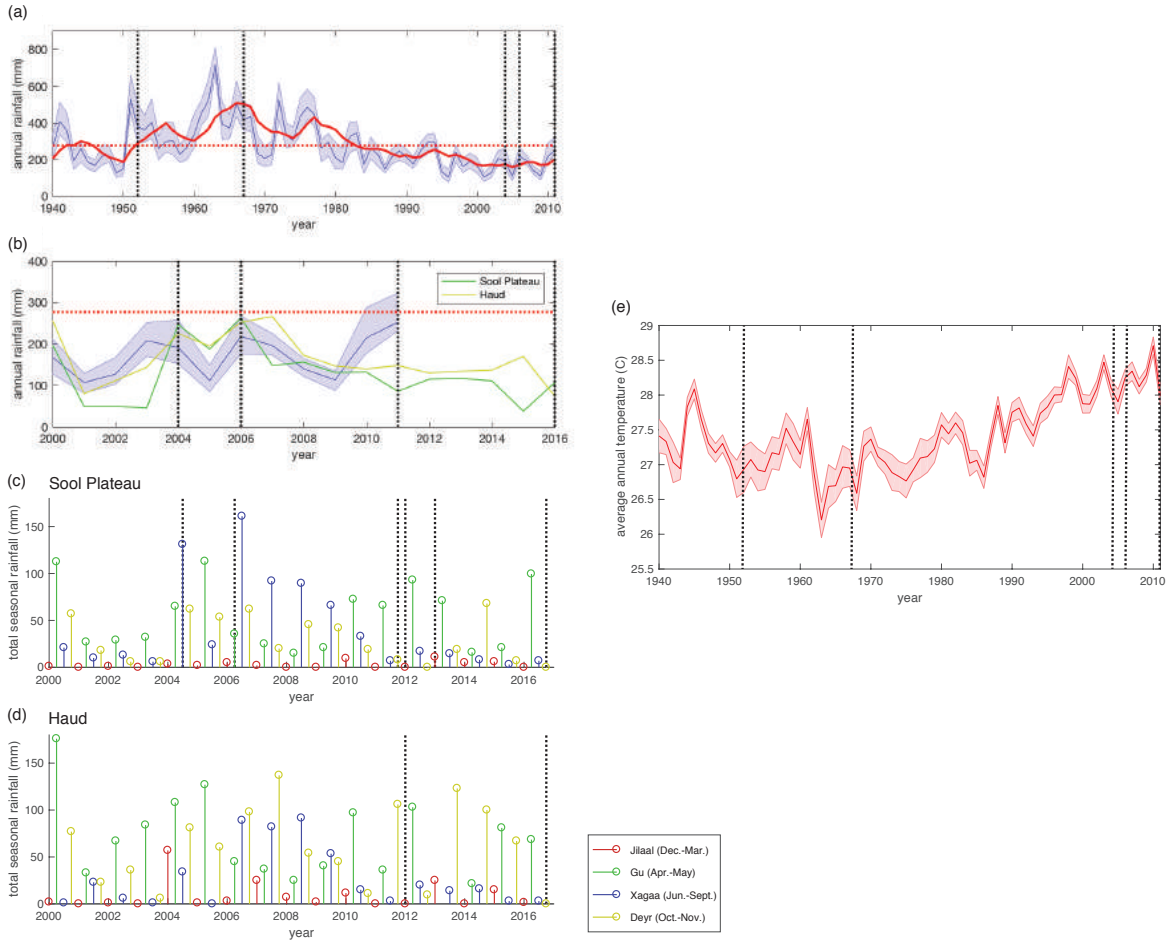

Figure S1: Annual total rainfall estimates from V2c and RFEv2 datasets, and temperature estimates from V2c dataset. (a) shows the median V2c annual total rainfall estimates between 1940 and 2011 for a large region which includes both Sool Plateau and Haud sites. The area between 25th and 75th percentiles is shaded. The running average of median precipitation over the previous 5 years is plotted with a red solid line. The average rainfall over the entire interval is indicated with a red dashed line. (b) shows V2c (2000-2011) and RFEv2 (2000-2016) annual rainfall datasets. The average V2c rainfall estimate over 1940-2011 is indicated with a red dashed line. (c) shows seasonal rainfall totals in an area containing the Sool Plateau study areas, and (d) shows totals in an area containing Haud study areas. (e) shows average yearly temperature (°C) computed from V2c reanalysis dataset. Average yearly temperature is defined as a yearly average over the daily midpoint between minimum and maximum temperatures. One standard deviation about the mean based on 56 reanalysis simulations is indicated with shading. Dates of imagery datasets are indicated in black dashed lines.

In recent decades, *Acacia bussei* has diminished in abundance in the Sool Plateau due to cutting for charcoal production, as it is considered the most lucrative species for this purpose [6]. Disruption of traditional grazing patterns has resulted in overgrazing in many areas of the Sool Plateau, including Dhahar (SP4).

### S1.3 Soil

Soils studied around areas with vegetation bands in the Haud [4] and Sool Plateau [6] are claylike and finely textured. In both regions soils appear prone to crust formation and soil pore plugging, resulting in low permeability and surface water runoff following high-intensity rainfall. Hemming found that soils are wetter beneath bands in the Haud, indicating greater soil permeability in vegetated areas [4]. Soils in the Sool

3

Plateau thinly cover a limestone bedrock, which is exposed in some parts of our study areas.

## S2 Data

### S2.1 Imagery

We studied approximately 260 km$^2$ of imagery in areas of the Sool Plateau and 200 km$^2$ of imagery in areas of the Haud. Study area boundaries are defined by our choice of British Royal Air Force (R.A.F.) aerial survey photography, which comprise our earliest image datasets. Aerial survey photographs were taken in 1951-52 over broad areas of British Somaliland, and are archived at the Bodleian Library at the University of Oxford. The aerial photographs used in this study were scanned on request by the Bodleian Library using British Ordnance Survey maps to identify images.

The coordinates of study areas and additional information about imagery used in this study are given in Table S1. R.A.F. images were scanned at a nominal resolution of 1.4-2.5 m/pixel. We obtained more recent imagery through the USGS and DigitalGlobe Foundation. We purchased declassified reconnaissance satellite imagery[1] taken in 1967 from the USGS Earth Explorer site. We downloaded freely available OrbView-3 imagery taken in 2005 from the USGS Earth Explorer site. We were granted QuickBird-2, WorldView-1, and WorldView-2 imagery for dates spanning 2004-2016 by the DigitalGlobe Foundation.

We manually georeferenced aerial survey photograph scans in ArcMap 10.3 against the ArcGIS World Imagery layer using the WGS84 Web Mercator coordinate system (EPSG:3857). Because vegetation bands migrate over time, we could not match scans with geospatial coordinates using the appearance of the bands themselves. Instead we relied upon apparent geological features, such as limestone outcrops, and geometrically distinct clusters of individual trees or shrubs that persisted over time. Aerial survey photographs were matched using no fewer than 10 control points per image, and were aligned by fitting a projective transformation. Control points were stored in a tab-delimited file. A projective transformation is overdetermined for greater than 4 control points, and the root mean squared error (RMSE) of the transformed control points served as our estimate of georeferencing error.

To estimate the effect on RMSE of adding additional control points, we used a resampling procedure that calculates the alignment RMSE for different subsets of the control points. For an image that was aligned using $n$ control points, we computed the RMSE for permutations of $5 \leq k \leq n$ control points. The average RMSE values over the permutations were then computed for each value of $k$. In this procedure, if the total number of such permutations $\binom{n}{k}$ exceeded $10^3$, a random sampling of $10^3$ distinct permutations were used. Otherwise, all permutations were used. The resulting curves were well fit by the saturating function $a\tilde{k}/(1+b\tilde{k})$, where $\tilde{k} = k - 5$, to extrapolate the saturating value of the average RMSE curve. In all cases the saturating RMSE value was comparable to the resolution of the imagery, suggesting an alignment error on the order of 1-2 pixels.

A reconnaissance satellite image was also manually georeferenced in ArcMap 10.3 using a third-order polynomial transformation with 18 control points. The image covers a much broader area than the aerial photographs, and due to distortions arising from the imaging methodology a projective transformation did not produce a suitable fit[2]. RMSE of this alignment is 0.94 m, which is on the order one pixel. DigitalGlobe imagery was pre-georeferenced, and precise alignment with the ArcGIS World Imagery layer required only manual translation.

Most recent satellite imagery used in this study contain data sensed at different frequency channels. The red, green, and blue channels were used for visualization, and the red and near infrared channels were used for computing the Soil-adjusted Vegetation Index (SAVI) [8], an index of photosynthetic activity:

$$\text{SAVI} = \frac{NIR - R}{NIR + R + L}(1 + L).$$

$NIR$ is the near-infrared reflectance value, and $R$ is the red reflectance value. We computed reflectances from raw pixel intensity values using radiometric calibration adjustment factors given by DigitalGlobe[3] and

---

[1] Corona program, Mission No. 1102-1
[2] A third-order polynomial fit is used for comparable imagery in [7]
[3] https://www.digitalglobe.com/resources/technical-information

| Area | (Lat, Lon) | Data area (km$^2$) | Date | Res. (m) | Channels used | Source | Sensor |
|---|---|---|---|---|---|---|---|
| SP1 | (9.79°, 48.55°) | 57 | 02/22/1952† | 1.9* | Grayscale scan | Bodleian | |
| | | | 12/12/1967 | 2.0* | Grayscale scan | USGS | KH-4B |
| | | | 06/10/2004 | 2.4 | R,G,B,NIR | DigitalGlobe | QuickBird-2 |
| | | | 03/23/2006 | 2.4 | R,G,B,NIR | DigitalGlobe | QuickBird-2 |
| | | | 09/29/2011† | 2.0 | R,G,B,NIR | DigitalGlobe | WorldView-2 |
| | | | 12/03/2011 | 2.0 | R,G,B,NIR | DigitalGlobe | WorldView-2 |
| | | | 02/24/2013 | 2.0 | Panchromatic | DigitalGlobe | WorldView-2 |
| SP2 | (9.72°, 48.55°) | 58 | 02/22/1952† | 1.9* | Grayscale scan | Bodleian | |
| | | | 12/12/1967 | 2.0* | Grayscale scan | USGS | KH-4B |
| | | | 06/10/2004 | 2.4 | R,G,B,NIR | DigitalGlobe | QuickBird-2 |
| | | | 03/23/2006 | 2.4 | R,G,B,NIR | DigitalGlobe | QuickBird-2 |
| | | | 09/29/2011† | 2.0 | R,G,B,NIR | DigitalGlobe | WorldView-2 |
| | | | 12/03/2011 | 2.0 | R,G,B,NIR | DigitalGlobe | WorldView-2 |
| | | | 02/24/2013 | 2.0 | Panchromatic | DigitalGlobe | WorldView-2 |
| SP3 | (9.60°, 48.59°) | 46 | 11/29/1952† | 1.4* | Grayscale scan | Bodleian | |
| | | | 12/12/1967 | 2.0* | Grayscale scan | USGS | KH-4B |
| | | | 06/10/2004 | 2.4 | R,G,B,NIR | DigitalGlobe | QuickBird-2 |
| | | | 03/23/2006 | 2.4 | R,G,B,NIR | DigitalGlobe | QuickBird-2 |
| | | | 12/03/2011† | 2.0 | R,G,B,NIR | DigitalGlobe | WorldView-2 |
| | | | 02/24/2013 | 2.0 | Panchromatic | DigitalGlobe | WorldView-2 |
| SP4 | (9.75°, 48.83°) | 58 | 02/22/1952† | 1.6* | Grayscale scan | Bodleian | |
| | | | 12/12/1967 | 2.0* | Grayscale scan | USGS | KH-4B |
| | | | 11/06/2005 | 1.0 | Panchromatic | USGS | OrbView-3 |
| | | | 08/16/2016† | 2.0 | R,G,B,NIR | DigitalGlobe | WorldView-2 |
| SP5 | (9.36°, 48.79°) | 44 | 02/14/1952† | 1.5* | Grayscale scan | Bodleian | |
| | | | 12/12/1967 | 2.0* | Grayscale scan | USGS | KH-4B |
| | | | 08/16/2016† | 2.0 | R,G,B,NIR | DigitalGlobe | WorldView-2 |
| HD1 | (8.14°, 47.21°) | 46 | 02/17/1952† | 2.5* | Grayscale scan | Bodleian | |
| | | | 11/24/2016† | 2.0 | R,G,B,NIR | DigitalGlobe | WorldView-2 |
| HD2 | (8.14°, 47.39°) | 50 | 02/14/1952† | 2.5* | Grayscale scan | Bodleian | |
| | | | 12/25/2011 | 2.0 | R,G,B,NIR | DigitalGlobe | WorldView-2 |
| | | | 01/21/2012† | 2.0 | R,G,B,NIR | DigitalGlobe | WorldView-2 |
| HD3 | (8.06°, 47.44°) | 50 | 01/24/1952† | 2.5* | Grayscale scan | Bodleian | |
| | | | 12/25/2011 | 2.0 | R,G,B,NIR | DigitalGlobe | WorldView-2 |
| HD4 | (8.09°, 47.47°) | 50 | 01/24/1952† | 2.5* | Grayscale scan | Bodleian | |
| | | | 12/25/2011† | 2.0 | R,G,B,NIR | DigitalGlobe | WorldView-2 |

Table S1: Study area locations and imagery datasets used in this study. Datasets used in visual comparison, Fourier analysis, and band statistics listed in Table 1 indicated with †. Nominal resolutions of photograph scans are indicated with *. R, G, and B denote red (630-690 nm for QuickBird-2 and WorldView-2), green (520-600 nm for QuickBird-2 and 510-580 for WorldView-2), and blue (450-520 nm for QuickBird-2 and 450-510 nm for WorldView-2) channels respectively. NIR denotes the near-infrared channel (760-900 nm for QuickBird-2 and 860-1040 nm for WorldView-2).

the imagery metadata. The parameter $L$ is used to adjust for exposed soil surface in low-vegetation cover scenarios, and is often used in place of Normalized Difference Vegetation Index (NDVI) in dryland vegetation inference. SAVI is equivalent to NDVI for $L = 0$. For visualizing SAVI values, we used a conventional parameter value of $L = 0.5$.

## S2.2 Elevation

We used NASA Shuttle Radar Topography Mission Global 1 arc second (SRTMGL1) elevation data for our upslope migration assessment and comparison of pattern properties with slope. Datasets were obtained from the USGS website[4]. Datasets are packaged in 1° latitude × 1° longitude tiles, and were projected onto the WGS84 Web Mercator coordinate system (EPSG:3857) in MATLAB 2016b. This allowed for elevation data to be matched with imagery.

Though the SRTMGL1 dataset has a nominal resolution of 1 arcsecond (∼30 m/pixel near the equator), the true resolution is closer to 45-60 m/pixel due to the manner in which data was collected [9]. The data also contains speckle which is autocorrelated at a length of 1-2 pixels, and also random error, both of which together result in average vertical error of approximately 4 m in areas like the Sahara Desert [9]. To eliminate autocorrelated errors, we subsampled the data to 3 arcsecond (∼90 m/pixel) resolution.

Speckle in the SRTM data presents a challenge to gradient estimation in areas of low relief, such as our regions of study, where in banded areas vertical change can be as little as 1 m per 500 m of horizontal change. To compute gradient fields, we used a second-order accuracy finite difference stencil with noise suppressing properties [10]. As an example, a $5 \times 3$ noise suppressing gradient operator as defined in [10] is

$$f = \frac{1}{32h} \begin{bmatrix} -1 & -2 & 0 & 2 & 1 \\ -2 & -4 & 0 & 4 & 2 \\ -1 & -2 & 0 & 2 & 1 \end{bmatrix},$$

where $h$ is the discretization step size. Convolving this operator with the data array produces an approximation of the partial derivative in one direction. Operators with noise suppressing properties discussed in [10] can be computed for arbitrarily large stencil size.

Using a finite difference operator allows for a straightforward propagation of independent and identically distributed normal errors in the elevation data through the calculation of gradient and slope. We estimated the magnitude of errors in the elevation data by computing the standard deviation of residuals from a median subtraction:

```
slidingmed = medfilt2(SRTM, [5 5]); %median-filtered data
sig = std(SRTM(:)-slidingmed(:)); %standard dev of residuals
```

The standard deviation of error propagated to each component of the gradient is then

```
eps = sig*sqrt(sum(f(:).^2))/h;
```

Slope is obtained from the magnitude of the elevation gradient vector, and to leading order the gradient error value is also equal to the error propagated to the slope calculation.

We tested the sensitivity of slope calculations to varying stencil size $s$ (which yields a $(2s+1) \times (2s-1)$ operator). We note again that the truncation error of the finite difference operator is second-order for any $s$. Intuitively too small a stencil size will have high measurement error, and too large a stencil size will result in oversmoothing. In Figure S2 we show the 25th, 50th, and 75th percentiles of the slope values within each study area computed over an interval of $s$, and indicate one standard deviation of propagated error around these values. We conclude that slope values are not sensitive to stencil size when $s \geq 15$, and we use 15 (which gives an operator of size $31 \times 29$) for all slope and gradient calculations. We confirmed by visual inspection that this stencil size produces smooth gradient fields that match hydrological features visible in the imagery (e.g., hills and channels).

## S3 Visual comparison

### S3.1 Protocol

We assessed changes over time at study areas via a systematic visual comparison of imagery. Roads can be visually identified in both the aerial photographs and the satellite imagery, and their presence and qualitative appearance served as our primary proxy for inferring the extent of human pressure. Vegetation in both the aerial photos and satellite imagery contrasts sharply with the light background of bare soil, and bands are

---
[4]https://e4ftl01.cr.usgs.gov/SRTM/

6

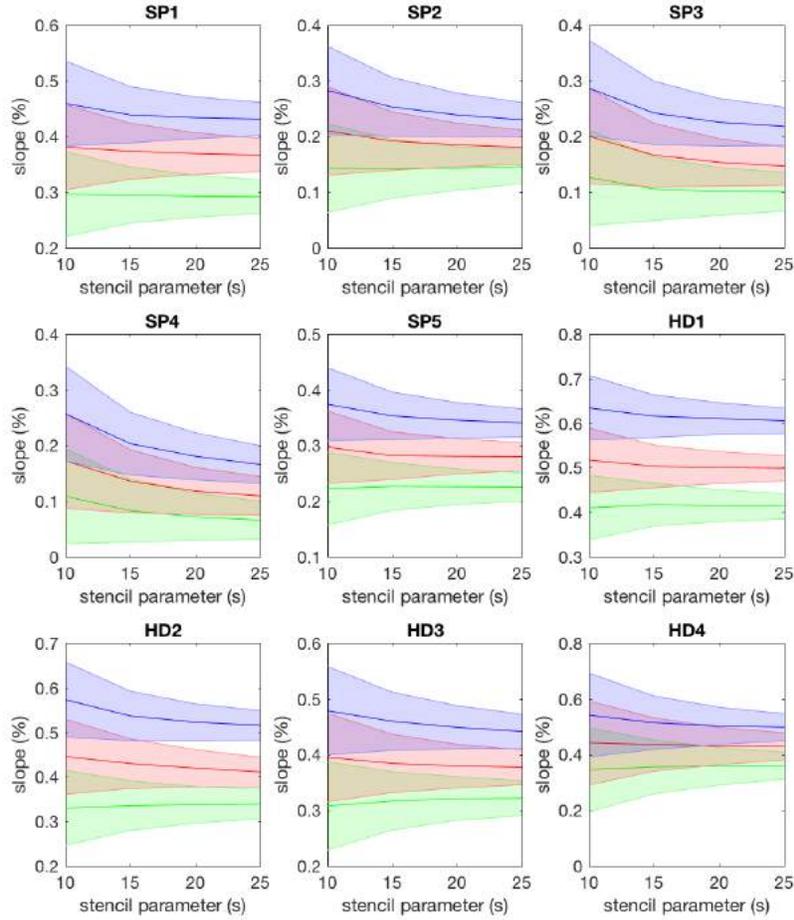

Figure S2: Sensitivity analysis of slope calculation to stencil parameter at different study areas. The 25th (green), 50th (red), and 75th (blue) percentiles of slope values are plotted as a function of the stencil parameter, and one standard deviation of propagated error are indicated in shading. Slopes are given in units of slope percentage, which is defined as 100 times the magnitude of the elevation gradient vector.

clearly identifiable. Degradation was inferred through either the breakdown in regularity or disappearance of banding.

We developed a graphical user interface (GUI) in MATLAB 2016b for visually comparing images (Figure S3). The GUI allows the user to select two imagery datasets for comparison, a georeferenced R.A.F. photograph and a more recent image. Images used for visual comparison are indicated in Table S1. The recent image is projected onto the intrinsic (row-column) coordinate system of the R.A.F. photograph, so that the data can be cleanly divided into non-overlapping 1 km × 1 km windows. The GUI simultaneously displays corresponding 1 km$^2$ windows of the R.A.F. photo and more recent imagery. Additionally the GUI displays a false color overlay of the two images, which was used to assess whether migration occurred. The GUI plots the local slope direction vector (computed as described in Section S2.2) on top of the overlay, which allows the user to visually assess whether the migration is in the upslope direction.

For each image window, the user is prompted to enter whether regular banding is present and whether a dense settlement is present (more than 5 structures in close proximity) using checkboxes. The user can select the extent of apparent road cover via a dropdown menu. Additionally, the user can check boxes to indicate whether band widening in the slope direction is apparent, and whether it appears that the same roads or

7

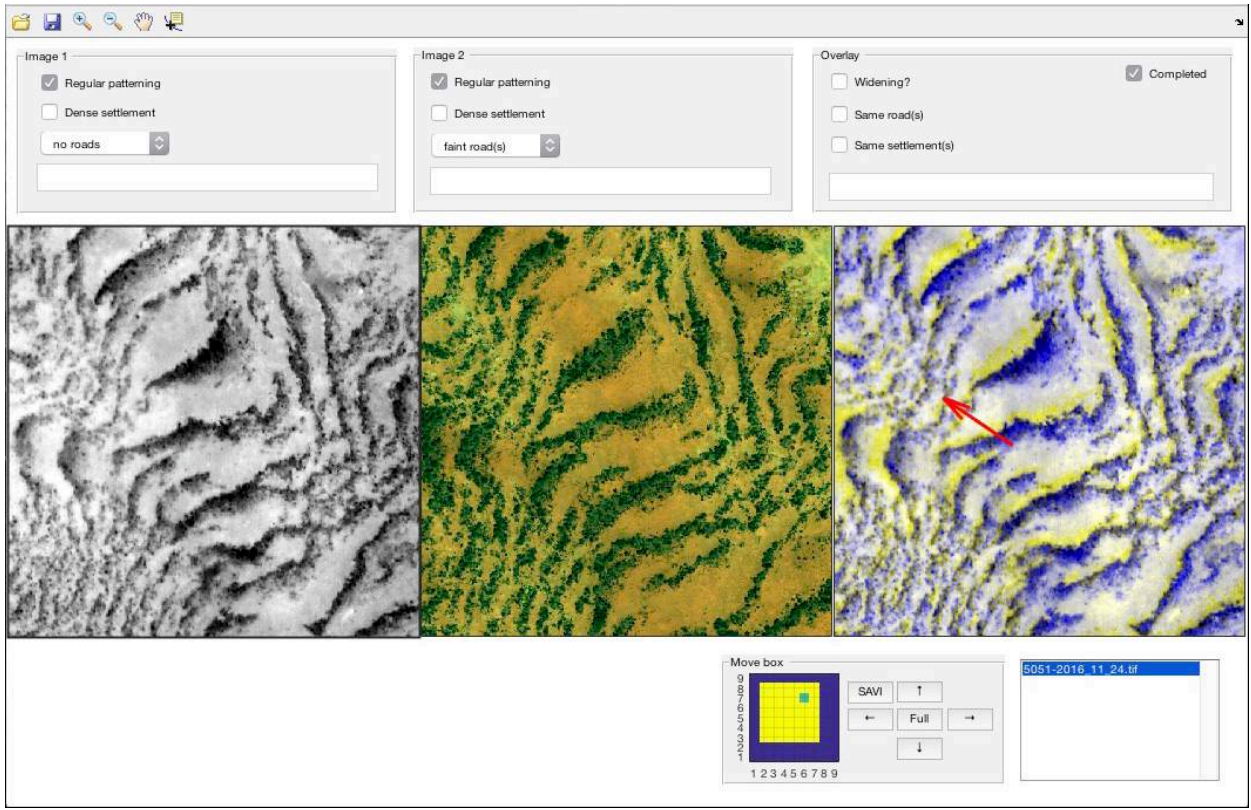

Figure S3: MATLAB GUI used for visual comparison of images. A georeferenced R.A.F. photograph and a recent satellite image are divided into corresponding windows and shown side by side (8.15° N, 47.23° E; 02/17/1952, 11/24/2016). A false color overlay of the two images is also shown, with a local slope vector overlaid to visually assess upslope migration. Blue in the false color image denotes vegetation in the R.A.F. photograph, and yellow denotes vegetation in the recent image. Images courtesy of the Bodleian Library and the DigitalGlobe Foundation.

settlements are present in both images. The user can enter comments for each image and the overlay. If the recent image contained red and near infrared channels, the Soil-adjusted Vegetation Index (SAVI) can be displayed in place of the RGB image (see discussion of SAVI in Section S2.1). The GUI selections are automatically saved to a MATLAB .mat file. When the user has finished assessing the window, the user can then navigate to different windows in the dataset using buttons.

The road cover dropdown can take on one of four states: "no roads," "faint road(s)," "clear road(s)," and "clear, dense road(s)." The last state, "clear, dense road(s)," is taken to mean that a large number of well-incised, clearly visible roads cover a large portion of the window. Examples of these classifications are shown in Figure S4. The checkboxes are ternary; with banding, for example, a fully-checked state is taken to mean distinct banding, a half-checked state is taken to mean indistinct banding, and an unchecked state is taken to mean no banding. Examples of these classifications are shown in Figure S5.

### S3.2 Highlighted examples

In many sites of the Haud study areas, we observed that human-made structures appeared to persist from 1952 to the present. In Figure S6a, we show examples of such structures.

In addition, we observed that in some areas of SP4, bands appeared to degrade without apparent change in wavelength. We show an example in Figure S6b. We verified that the bands shown have significantly lower vegetation index (SAVI) values than nearby bands in the study area, and are plausibly degraded.

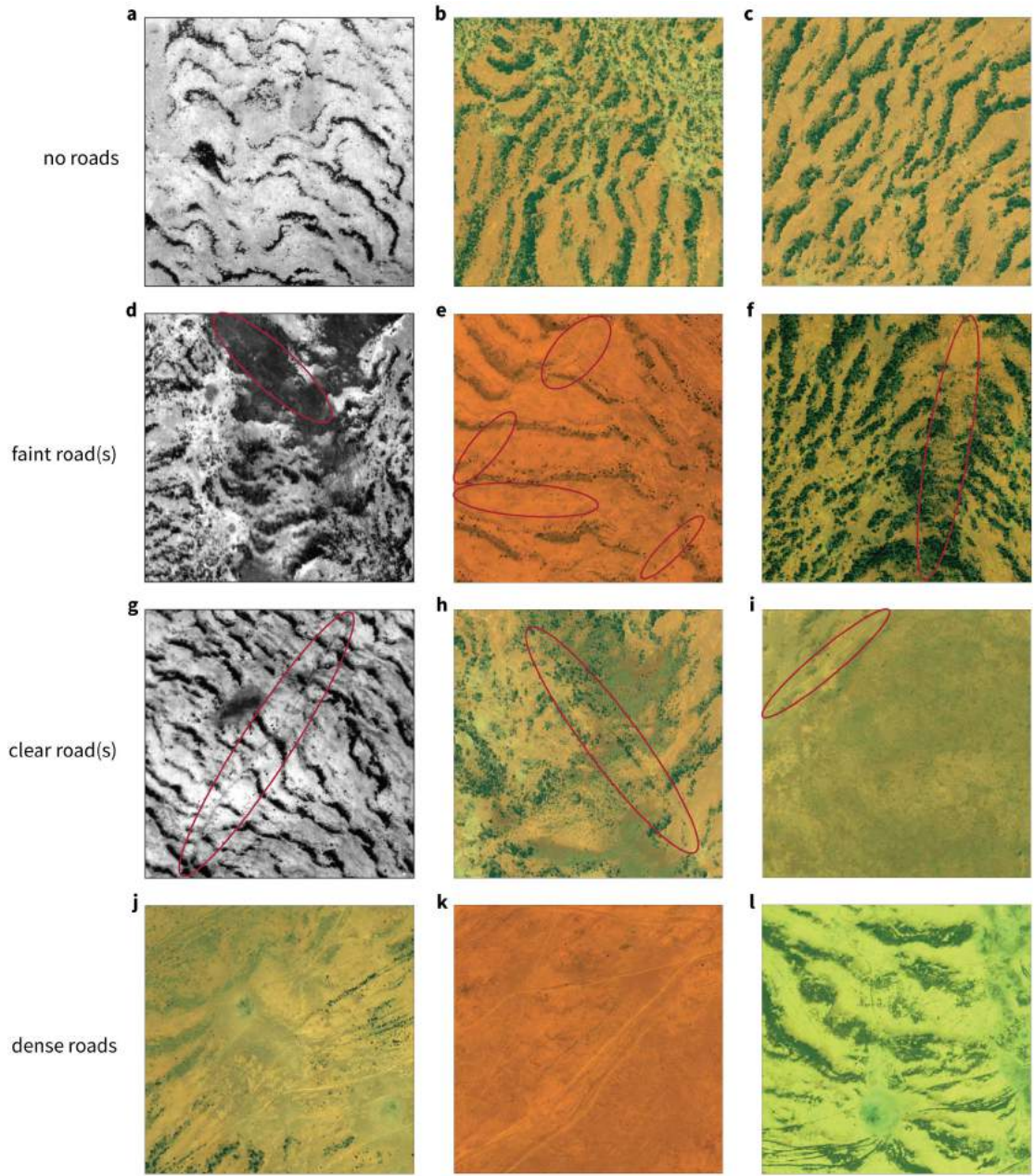

Figure S4: Examples of visual classifications of road cover state. (a)-(c) show examples of a "no roads" classification (a: SP3, 11/29/1952, grayscale; b: HD3, 12/25/2011, RGB; c: HD4, 12/25/2011, RGB). (d)-(f) show examples of a "faint road(s)" classification, where observed roads are circled (d: HD4, grayscale; e: SP5, 08/16/2016, RGB; f: HD1, 11/24/2016, RGB). (g)-(i) show examples of a "clear road(s)" classification (g: SP5, 02/14/1952, grayscale; h: HD4, 12/25/2011, RGB; i: SP3, 12/03/2011, RGB). (j)-(l) show examples of a "dense roads" classification (j: SP3, 12/03/2011, RGB; k: SP4, 08/16/2016, RGB; l: SP2, 09/29/2011, RGB). Colors shown for RGB images are normalized reflectance values computed from raw pixel intensities. Images courtesy of the Bodleian Library and the DigitalGlobe Foundation.

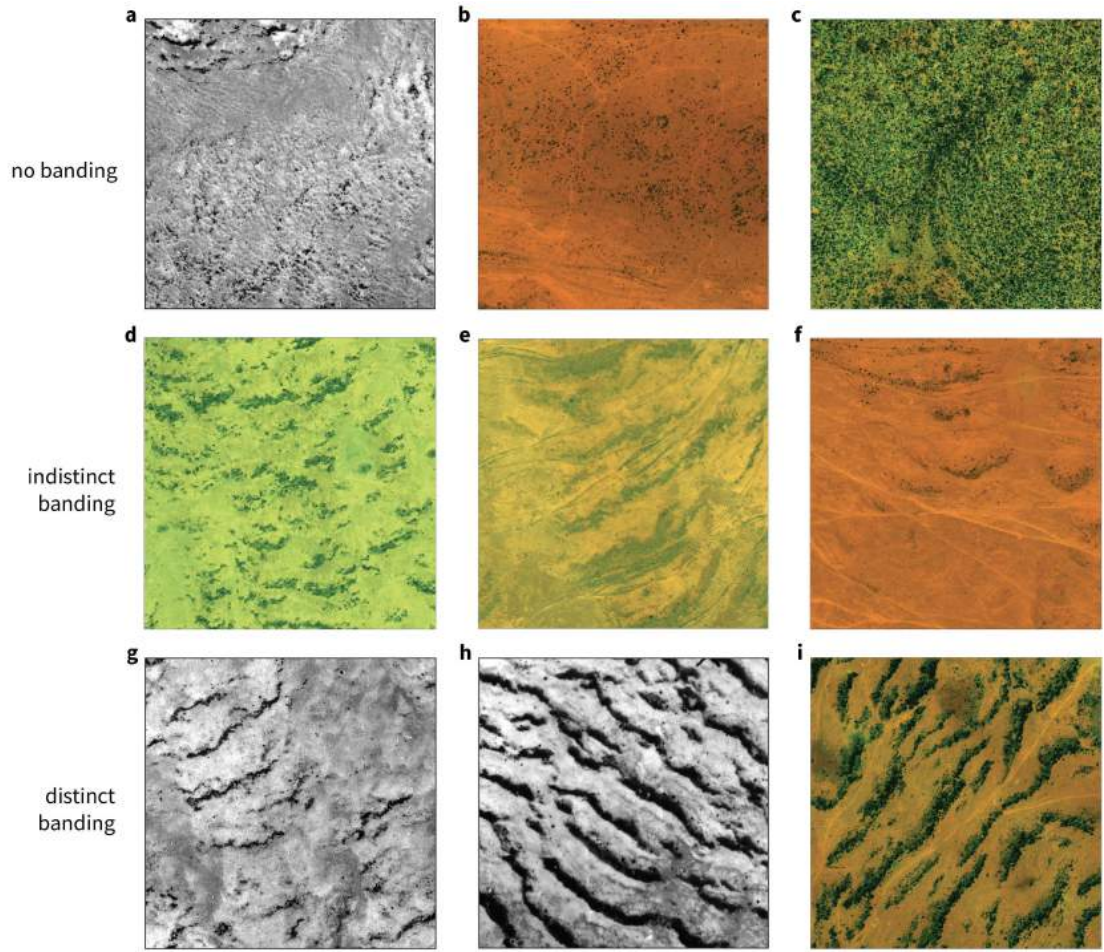

Figure S5: Examples of visual classifications of banding state. (a)-(c) show examples of a "no banding" classification (a: SP3, 11/29/1952, grayscale; b: SP4, 08/16/2016, RGB; c: HD2, 01/21/2012, RGB). (d)-(f) show examples of an "indistinct bands" classification (d: SP1, 09/29/2011, RGB; e: SP3, 12/03/2011, RGB; f: SP4, 08/16/2016, RGB). (g)-(i) show examples of a "distinct bands" classification (g: SP4, 02/22/1952, grayscale; h: SP5, 08/16/2016, grayscale; i: HD1, 11/24/2016, RGB). Colors shown for RGB images are normalized reflectance values computed from raw pixel intensities. Images courtesy of the Bodleian Library and the DigitalGlobe Foundation.

## S4 Automated transect measurements

### S4.1 Protocol

We quantified aspects of vegetation dynamics using automated transect measurements of individual bands. To do this, we segmented the aerial photograph to identify bands, gathered image intensity profiles along transects through the bands in direction of slope, and fit a simple top hat function to extract band width and position along the transect. We used this information to assess changes in band width over time, as well as band migration.

We eliminated the large-scale background variations in pixel intensity in the aerial photographs by subtracting a coarsely Gaussian-blurred version of the image. We then applied a manually-tuned threshold to create a binary image of the vegetation bands. We passed this binary image to the `regionprops` function in MATLAB 2016b, and extracted the areas and centroids of connected components, as well as the major and minor axis lengths and orientations of ellipses fit to the connected components. We applied manually tuned criteria on the areas and ratio between major and minor axis length to isolate the vegetation bands in the binary image.

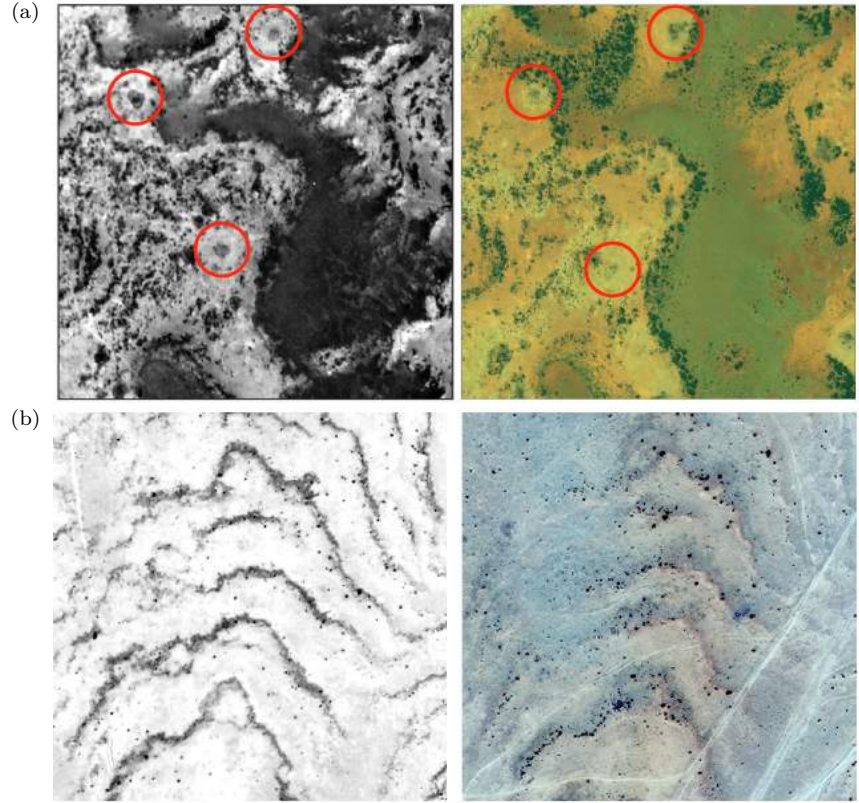

Figure S6: Highlighted examples from visual inspection. (a) shows an example at HD4 of human-made structures that appear to persist from 1952 to 2011 (8.08° N, 47.48° E; 01/24/1952, 12/25/2011). (b) shows an example at SP4 where degradation appears to have occurred without an apparent history of change in band wavelength (9.78° N, 48.84° E; 02/22/1952, 08/16/2016). Images courtesy of the Bodleian Library and the DigitalGlobe Foundation.

We then drew linear transects through the centroids of the bands in the direction of the minor axis. We visually confirmed that the minor axis direction serves as an effective proxy for the slope direction. For all study areas, transect lengths are approximately 100 pixels (∼190 m). The transects were drawn so that 25% of the transect lies downslope of the centroid, and 75% lies upslope. We did this so that the same transect could be used for both the aerial photographs and the more recent imagery, accounting for band migration upslope. In order to obtain replicate measurements for estimation of variance, we drew eight additional transects transverse to the original (four on either side). These transects are spaced approximately 4 m apart, which precludes double-sampling of pixels by adjacent transects. We then used these transects to extract pixel intensity profiles of the aerial photography and more recent imagery. We converted color images to grayscale before extracting intensities using the `rgb2gray` MATLAB function.

We fit the intensity profile with simple plateau-like curves using MATLAB's nonlinear least squares curve fitting function, `lsqcurvefit`. The curve has the form

$$f(x; \mathbf{b}) = b_1 + \frac{b_2}{2}\left[\tanh(\alpha(x - b_3)) - \tanh(\alpha(x - b_4))\right],$$

which approaches a piecewise constant function with levels $b_1$ and $b_2$ and breakpoints at $b_3$ and $b_4$ in the limit as $\alpha \to \infty$ (Figure S7). We used $\alpha = 500$. To fit this function, we rescaled all transects to lie along the interval $x \in [0, 1]$. For our data, the squared error cost function typically had many local minima, and so the result was sensitive to the initial guess for the parameters $b_3$ and $b_4$. We fit each intensity profile using 20 uniformly random initial guesses for $b_3$ and $b_4$ (such that $b_3 < b_4$), and chose among these the result with the minimum squared error. We then used the value $w = b_4 - b_3$ as the measured width of the band along the particular transect, and compared $b_3$ and $b_4$ values along a transect at different time points to measure migration.

11

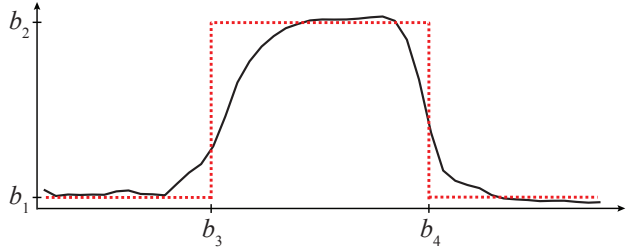

Figure S7: Schematic example of pixel intensity profile along a transect (black) and plateau function fit to the profile (red). Levels $b_1$ and $b_2$, as well as breakpoints $b_3$ and $b_4$ are indicated.

Since each band was measured using multiple parallel transects, we obtained multiple measurements of $w$, $b_3$, and $b_4$ for each band. We used a threshold on the standard deviation ($\sigma \leq 0.2$) of the $w$ measurements to exclude data points where bands may have substantially degraded or disappeared, or where the measurement is likely poor for some other reason. After applying the threshold, we calculated the mean $w$ for each remaining band at each time point.

### S4.2 Sool Plateau measurements

We observed appreciable increases in band width in SP1-SP4. We measured widths at multiple time points in these areas to assess when the widening may have occurred and whether it is a seasonal phenomenon. In Figure S8, we plot the distribution of band widths in the Sool Plateau sites SP1-SP5 over time. We have reconnaissance imagery taken in 1967 for all these sites. We observe that band widths changed little between 1952 and 1967. In SP1–SP4, widths are then larger in the recent imagery (onward from 2004), and do not return to their 1952/1967 widths. In SP5, widths remain unchanged between 1967 and 2016. We conclude that band widths increased in SP1–SP4 sometime between 1967 and 2004, and that this widening is not a seasonal effect.

## S5 Fourier analysis

### S5.1 Protocol

We quantitatively assessed changes in band wavelength using a modification of the Fourier window method by Penny *et al.* [11]. Penny *et al.* developed the method to compute spatial maps of local wavelength and orientation from imagery over banded areas in Fort Stockton, Texas, USA. In a manner analogous to a short-time Fourier transform, the method measures wavelength and orientation in a sliding window using a 2D FFT. Vegetation banding typically contains sufficient irregularity to complicate the inference of dominant wavelength and orientation from a 2D power spectrum. The Fourier window method addresses this issue by binning power, radially for estimating wavelength and angularly for orientation, and by computing a weighted average among the contiguous bins with largest power. The method computes a uniqueness metric for both wavelength and orientation based on the distance between the maximal peak and the nearest peak with 75% of the maximal power, if present. The metric equals one if the maximal peak is the only powerful peak present, and approaches 0 as distance to the nearest powerful peak increases. In order to exclude short-wavelength noise and long wavelengths which are under-sampled for the given window size, the bins are only computed for a specified minimum and maximum wavelength interval. As presented in [11], the pattern irregularity issue is also addressed by averaging measurements over overlapping windows. We do not perform the latter step for the analyses in this study.

Penny *et al.* [11] provide MATLAB code for their method, which we modify for our analyses. We modify the main routine to take as input two images that have been resized to the same dimensions. Computations are then performed on square windows. To reduce aperiodicity effects, we apply a 2D Hamming filter (a bell-shaped function that decays to zero away from the center) to each window. We note that this filter

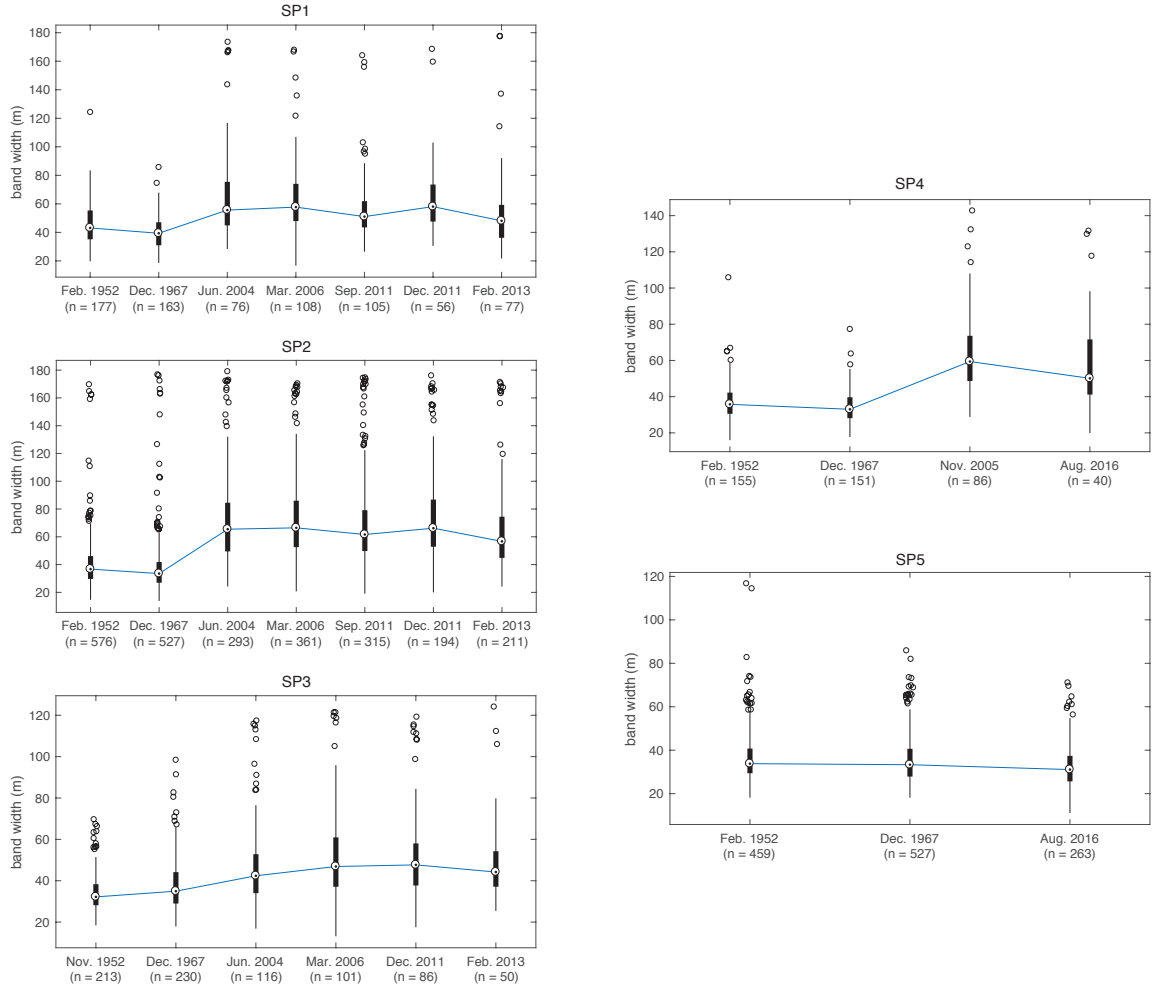

Figure S8: Band widths measured at SP1–SP5 are shown at multiple points in time. At SP1–SP4, widths change little between 1952 and 1967, and increase between 1967 and recent imagery. Widths remain approximately constant at SP5 from 1952-2016.

also has the effect of giving more weight to the central area of the image, focusing analysis on this area. Wavelengths and orientations are then computed using Penny *et al.*'s routine.

We applied this methodology to all study areas using the image pairs indicated in Table S1. In order to perform the windowing on a rectangularly-oriented dataset, we transformed the recent imagery onto the intrinsic coordinate system of the aerial photograph. We downscaled image pairs to a resolution of approximately 2.5 m/pixel to reduce computation time. We then applied two layers of preprocessing to emphasize the vegetation bands and de-emphasize other features in the imagery: we subtracted a coarsely gaussian-blurred version of the image to eliminate large scale variations in pixel intensity (such as darkening near the borders of the aerial photographs), and we applied a manually-tuned threshold to create a binary image of the vegetation bands.

For each binary image pair, we computed wavelength maps for the three square window sizes: 384 pixels (∼1 km), 512 pixels (∼1.3 km), and 768 pixels (∼2 km). After applying a Hamming filter, about 4-8 vegetation bands can be sampled in the central area of a 512 pixel window (3-5 bands for a 384 pixel window, or 6-10 for a 768 pixel window). For the binning procedure, we set the minimum wavelength to 10 pixels (∼25 m), and the maximum wavelength to one-fourth of the window size (∼240 m for a 384 pixel window, ∼320 m for a 512 pixel window, and ∼480 m for a 768 pixel window). In order to balance computation time

and even-sampling of the data, we set the step length of the sliding window to be one-fourth of the window size, resulting in adjacent windows that overlap in 75% of their area.

After wavelength maps were computed for an imagery pair, we transformed the measurements from units of pixels to units of meters. We then manually drew a mask on the imagery and applied it to the measurements in order to exclude measurements from areas without vegetation bands. Additionally we excluded measurements with wavelength uniqueness metrics smaller than 0.75. We found that in some areas, a window size of 384 pixels was too small to detect the largest wavelengths. The results of 512 and 768 pixel windows did not differ strongly, so we used the 512 pixel window computation for the results reported in the Table S2.

| Area | Slope (%) $S$ | Wavelength (m) $W_1$ | $W_2$ | WL change $W_2/W_1 - 1$ | Slope-WL corr. $\mathrm{corr}(S, W_1)$ ($p$, $t$, df) |
|---|---|---|---|---|---|
| SP1 | 0.3–0.4 | 130–170 | 130–190 | 0–10% | -0.15 (0.29, 1.3, 51) |
| SP2 | 0.1–0.3 | 130–170 | 140–170 | 0–10% | 0.02 (0.89, 0.0, 83) |
| SP3 | 0.1–0.3 | 120–150 | 140–180 | 0–20% | -0.34 (0.04, 4.5, 35) |
| SP4 | 0.1–0.2 | 130–160 | 150–180 | 0–20% | 0.08 (0.53, 0.4, 67) |
| SP5 | 0.2–0.4 | 120–140 | 120–140 | 0–10% | -0.23 (0.06, 3.7, 63) |
| HD1 | 0.4–0.6 | 80–100 | 80–120 | 0–10% | -0.25 (0.33, 1.0, 16) |
| HD2 | 0.3–0.5 | 90–110 | 90–120 | 0–10% | -0.19 (0.20, 1.7, 45) |
| HD3 | 0.3–0.5 | 80–100 | 80–110 | 0–10% | -0.14 (0.17, 1.9, 94) |
| HD4 | 0.4–0.5 | 100–120 | 100-120 | 0–10% | 0.08 (0.58, 0.3, 47) |

Table S2: Band properties measured using a modification of the Fourier window method by Penny *et al.* [11]. Ranges shown are the 25th and 75th percentiles. Wavelengths $W_1$ were measured in the R.A.F. aerial photography datasets, and $W_2$ were measured in recent satellite imagery datasets. Slopes were computed from the SRTM 1 arc-second elevation dataset. Significant negative correlation highlighted red. Significance of correlations was assessed using a *t*-test corrected for spatial autocorrelation [12], and *p* values, *t* values and degrees of freedom are given in parentheses.

### S5.2 Wavelength change

Given spatial maps of wavelength for a pair of images, we computed change maps where elements are given by

$$W_2^{i,j}/W_1^{i,j} - 1, \tag{1}$$

where $W_1^{i,j}$ is the wavelength in the first image at position $(i, j)$, and $W_2^{i,j}$ is the wavelength in the second image at $(i, j)$ (Figure S9). Typical ranges of the computed changes at each study area are given in Table S2. Note that for computing change maps, we have manually masked out areas with no banding, and we have also excluded measurements with a wavelength uniqueness metric smaller than 0.75. We made the latter choice to reduce the incidence of falsely detected changes between the maps, reasoning that measurements in areas with multiple dominant band wavelengths are error prone. Typical change ranges between 0–10% for all study areas except SP3 and SP4, where change ranges between 0–20%.

Our previous visual inspection suggested that there were no obvious systematic changes in wavelength at any study area, except perhaps for those associated with isolated instances of band loss in human-impacted areas. We visually reinspected areas where measured change was greater than 25% in magnitude. In some cases, it appears that these detected changes occur due to the loss of an individual band, often near evidence of human activity (Figure S10). In Figure S10b-e, significant road cover appears in the interband areas of the recent imagery, and are likely related to the loss of bands in these areas. In most cases, however, we saw no clear indication for the detected wavelength changes, and attributed these false detections to wavelength measurement error that arises due to the irregularity of the banding.

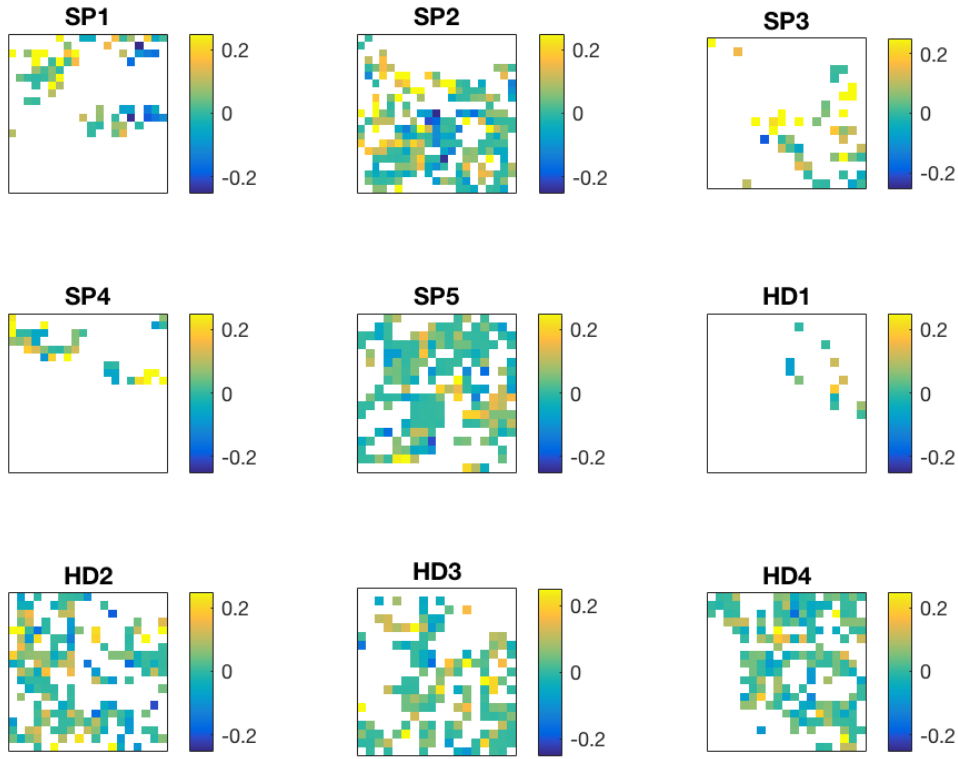

Figure S9: Wavelength change maps for all study areas. Wavelength change defined in (1), and computed between image pairs indicated in Table S1. White pixels indicate data points in areas without banding or areas with sufficiently low values of the computed wavelength uniqueness metric. All color axes are scaled to the interval [-0.25,0.25].

### S5.3 Wavelength-slope correlations

We computed the correlation between local wavelength and slope, and the results are reported in Table S2). We used slope values that are closest to the center point of the window corresponding to the wavelength/migration measurement. To assess the significance of correlations, we used a paired t-test for which sample size is corrected to account for spatial autocorrelation in the data [12]. The test is implemented in the library SpatialPack for R [13]. The correlation between wavelength and slope has been empirically investigated in [11] and [7]. We found no significant correlation between slope and wavelength for our study areas except for SP3 ($r = -0.34, p = 0.04$).

## S6 Model simulation

To explore how bands widen in response to parameter variation in a conceptual partial differential equation vegetation model, we simulated the model by Klausmeier [14] in one spatial dimension:

$$\begin{aligned}
N_T &= -MN + JRWN^2 + D_N N_{XX}, \\
W_T &= A - LW - RWN^2 + VW_X.
\end{aligned} \tag{K99}$$

Descriptions, units, and values of the parameters used are given in Table S3. The parameter set used for K99 is based on the values given in [14]. Parameters which are stated in [14] to differ between grasses and trees ($M$, $J$, and $R$) are set at intermediate values so that the spatial scale of banding resembles the scales in our regions of study. Water flow rate $V$ was also approximately tuned so that a comparable time

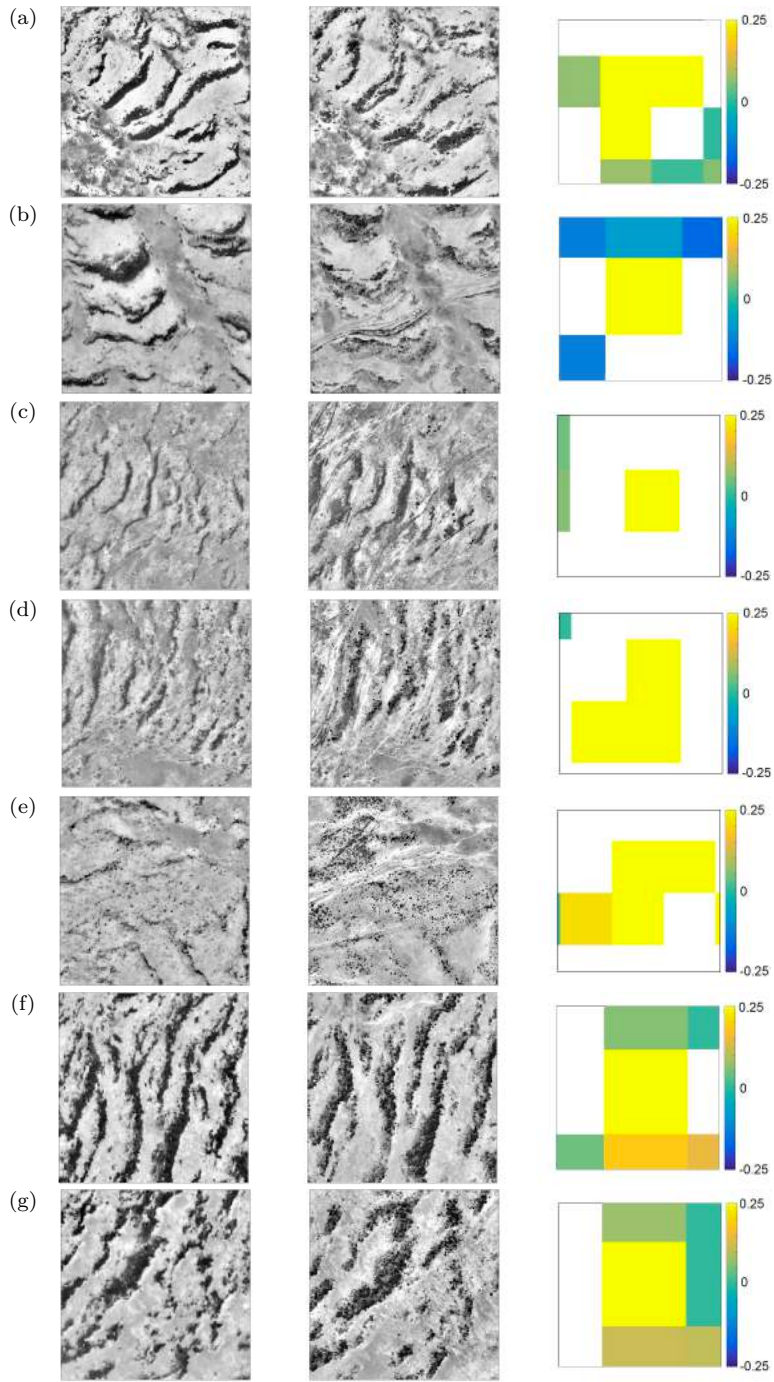

Figure S10: Areas where detected wavelength change corresponds to band loss or degradation. The first column is R.A.F. aerial photography, the second column is recent satellite imagery, and the third column is the detected wavelength change map. Locations of examples are as follows: (a) SP1 (9.77° N, 48.57° E; 02/22/1952, 09/29/2011), (b) SP2 (9.72° N, 48.54° E; 02/22/1952, 09/29/2011), (c) SP3 (9.60° N, 48.60° E; 11/29/1952, 12/03/2011), (d) SP3 (9.61° N, 48.61° E; 11/29/1952, 12/03/2011), (e) SP4 (9.76° N, 48.85° E; 02/22/1952, 08/16/2016), (f) HD4 (8.07° N, 47.46° E; 01/24/1952, 12/25/2011), and (g) HD4 (8.12° N, 47.46° E; 01/24/1952, 12/25/2011). Images courtesy of the Bodleian Library and the DigitalGlobe Foundation.

16

| Parameter/variable | Units | Description | Value |
|---|---|---|---|
| $A$ | mm H$_2$O yr$^{-1}$ | mean annual rainfall | 150 |
| $L$ | yr$^{-1}$ | evaporation rate | 4 |
| $J$ | kg m$^{-2}$ (mm H$_2$O)$^{-1}$ | biomass yield per unit H$_2$O | 0.0025 |
| $M$ | yr$^{-1}$ | mortality rate | 0.75 |
| $R$ | mm H$_2$O yr$^{-1}$ (kg dry mass)$^{-2}$ | transpiration rate | 50 |
| $D_N$ | m$^2$ yr$^{-1}$ | plant dispersal rate | 1 |
| $V$ | m yr$^{-1}$ | water flow speed | 35 |
| $N$ | kg dry mass m$^{-2}$ | plant biomass | |
| $W$ | mm H$_2$O (or kg H$_2$O m$^{-2}$) | water | |
| $X$ | m | spatial dimension along | |
| $T$ | yr | time | |

Table S3: Parameters and variables for the model by Klausmeier [14].

scale of migration is reproduced in the simulations to that of our regions of study. We set the mean annual rainfall parameter to 150 mm, which is within the range of typical rainfall levels in our regions of study. We simulated K99 using the ETDRK4 explicit pseudospectral scheme [15], with 2048 grid points, a 1000 m domain, and a time step of 0.01 years.

We performed a sensitivity analysis to estimate the linear response of band width and peak band biomass to changes in the parameter set. We began each simulation using the parameter set shown in Table S3 with an initial state of the uniform equilibrium value plus small-magnitude spatial noise. We evolved the initial state to 10,000 years to obtain an equilibrium migrating patterned state. We then began a set of perturbation simulations, where in each we perturb one value in the parameter set listed in Table S3 by a percentage between 5 and 100% that is manually tuned to produce a 5-10% response in width ratio. We then evolve the initial equilibrium patterned state by 50 years. The resulting patterned states are pulselike (Figure S11a), and we measured widths by thresholding using a small value ($10^{-3}$). We show the width ratios for all parameter perturbation simulations in Figure S11b, where the ratios are computed by dividing band widths in the perturbed simulations by the widths from the initial patterned state. Increasing $A$, $J$, $R$, $D_N$, and $V$ and decreasing $L$, and $M$ results in band width increases.

To simulate a scenario where vegetation species composition shifts from woody to grass biomass, we simultaneously increase $J$, $R$, and $M$ by 10% and $D_N$ by 50% (Figure S11b). Although increasing mortality by itself reduces the band width, the simultaneous increase of these four parameters results in band width increase.

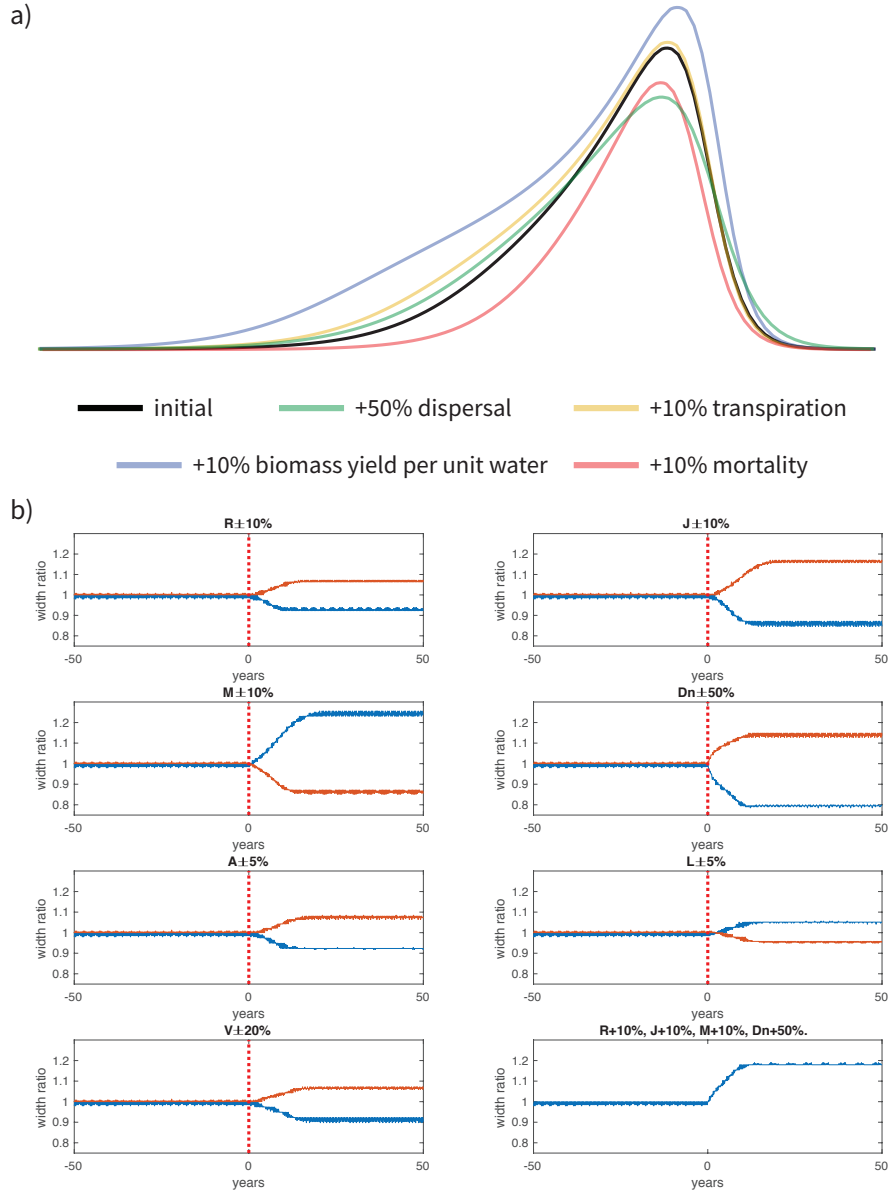

Figure S11: K99 sensitivity analysis to width ratio. (a) A comparison of equilibrium band profiles, simulated first using an initial parameter set, and then simulated after applying perturbations to individual parameters. (b) Parameters are perturbed individually by + (red) or - (blue) the percentage of the parameter indicated.



\end{document}